%% file: main.tex
\documentclass[a4paper]{scrartcl}
\usepackage[margin=1in]{geometry}
\usepackage[utf8]{inputenc}
\usepackage[english]{babel}
\usepackage{csquotes}
\usepackage{pgfkeys,pgfcalendar} 
\usepackage[pdftex]{hyperref}
\usepackage{mdframed}
\usepackage{todonotes}
\usepackage{multicol}
\usepackage{caption}
\usepackage{subcaption} % subfigures
\usepackage{longtable}
\usepackage{booktabs}
\usepackage{colortbl}
\usepackage{xcolor}

\usepackage[backend=biber, style=apa]{biblatex}
\title{The UN Security Council debates\\1992-2023}
\author{Mirco Schoenfeld\thanks{mirco.schoenfeld@tum.de} \and Steffen Eckhard\thanks{steffen.eckhard@zu.de} \and Ronny Patz\thanks{science@ronny-patz.de} \and Hilde van Meegdenburg\thanks{h.van.meegdenburg@fsw.leidenuniv.nl} \and Antonio Pires\thanks{antonio.hps26@gmail.com}}
\date{\pgfcalendarmonthname{\month} \number\year}
\AtBeginDocument{
  \hypersetup{
    pdftitle = {The UN Security Council debates 1995-2023},
    pdfauthor = {Mirco Schoenfeld, Steffen Eckhard, Ronny Patz, Hilde van Meegdenburg, Antonio Pires},
    pdfcreator = {ms},
    pdfproducer = {ms}
  }
}

\newcommand\numspeeches{106,302}
\newcommand\numspeecheslastyear{7,621}
\newcommand\nummeetings{6,233}
\newcommand\nummeetingslastyear{271}
\newcommand\coveredperiodend{2023}

\begin{document}

\begin{titlepage}
\maketitle
\vspace{-5mm}
\textbf{Abstract}\\[2mm]
This paper presents an updated dataset containing \numspeeches{} speeches held in the public meetings of the UN Security Council (UNSC) between 1992 and \coveredperiodend{}. The dataset is based on publicly available meeting transcripts with the S/PV document symbol and includes the full substance of individual speeches as well as automatically extracted and manually corrected metadata on the speaker, the position of the speech in the sequence of speeches of a meeting, and the date of the speech. After contextualizing the dataset in recent research on the UNSC, the paper presents descriptive statistics on UNSC meetings and speeches that characterize the period covered by the dataset. Data highlight the extensive presence of the UN bureaucracy in UNSC meetings as well as an emerging trend towards more lengthy open UNSC debates. These open debates cover key issues that have emerged only during the period that is covered by the dataset, for example the debates relating to Women, Peace and Security or Climate-related Disasters.\\%[2mm]

\textbf{Corpus}\\[2mm]
The UN Security Council Debates corpus is available online at \url{https://doi.org/10.7910/DVN/KGVSYH        }\\%[2mm]

\textbf{Citation}
\begin{mdframed}
Schönfeld, M., Eckhard, S., Patz, R., van Meegdenburg, H. \& Pires, A. (2025). The UN Security Council Debates 1995-2023. Retrieved from: https://arxiv.org/abs/1906.10969.
\end{mdframed}
\vspace{5mm}

\textbf{Acknowledgements}\\[2mm]
This paper presents an updated version of the publication "Schönfeld, M., Eckhard, S., Patz, R., \& Van Meegdenburg, H. (2019). The UN Security Council debates 1995-2017. arXiv preprint arXiv:1906.10969". In 2021, Antonio Pires joined the author team and was mostly responsible for the updated version of the dataset (years 2018-2023). New features include detailed information on agenda items based on the Repertoire of the Practice of the Security Council and variables that identify whether a speech was delivered in person or via video-teleconference (VTC).

\end{titlepage}

\section{Introduction}
The creation of the United Nations (UN) on the ruins of the League of Nations has profoundly shaped the post-World War II global order. At the center of the UN system with its myriad of agencies, committees, and conferences in New York, Geneva, Nairobi or Vienna thrones the most visible of UN institutions and key body in the global security architecture: the United Nations Security Council (UNSC). 

After intensive debates ahead of its creation \parencite[10-38]{2009-bosco-five}, the UNSC met for the first time on January 17, 1946 \parencite[41]{2009-bosco-five}. Initially composed of five permanent members (the P5) and six elected members, the UNSC since 1966 has 15 member countries, ten of which are non-permanent and represent the five regional groupings of the UN.\footnote{These are: African Group; Asia and the Pacific Group; Eastern European Group; Latin American and Caribbean Group; Western European and Others Group} The UNSC meets regularly in a public format allowing for the global public to follow key debates and votes. Although public meetings  are only the visible tip of the iceberg, as meetings are held under various formal and informal procedures \parencite{2014-sievers-unsc_procedure}, these meetings are among the best reported regular activities of the United Nations. This is particularly true when the Council meets on emerging crises or on matters of significant geopolitical tensions.
%, but public meetings are only the visible tip of the iceberg of meetings held under various formal and informal procedures \parencite{2014-sievers-unsc_procedure}. Nevertheless, the public meetings of the UNSC are among the best reported regular activities of the United Nations, in particular when the Council meets on emerging crises or on matters of significant geopolitical tensions. 
Studying the substance of these debates therefore gives insights into developments of historical importance. 
The emergence of new themes and topics indicates that
%when 
the world is changing 
and that
%or how 
states address new global challenges such as climate change.
%global warming. 
Furthermore, shifts in attention to particular topics or conflicts discussed in the Council can reveal emerging power constellations. And the shifts in tone of speeches and in the intensity of blame exchanged between speakers can reflect growing or declining geopolitical tensions. Thus, even when major decisions are taken behind closed doors, the language used in
%for
the public documents captures
%well
the dynamics of international politics as well as the evolution of multilateralism well.
%The emergence of new topics indicates when the world is changing: The change in attention to particular aspects of each conflict discussed in the Council can reveal emerging power constellations, while the shifts in tone of speeches and in the intensity of blame exchanged between speakers can reflect growing or declining geopolitical tensions. Thus, even when major decisions are taken behind closed doors, the language used for the public should

The dataset presented in this paper \parencite{2019_unsc_corpus}\footnote{The data is available online at \url{https://doi.org/10.7910/DVN/KGVSYH}} allows to trace all %these 
public UNSC debates over a 32-year period from 1992 until \coveredperiodend{} through qualitative and quantitative text analysis. As the unit of analysis, the dataset includes each individual speech contribution made by a participant of a public UNSC meeting. Overall, the data set covers \nummeetings{} public meetings and includes \numspeeches{} individual speech contributions.

Earlier and preliminary version of the dataset have been used or cited to study a wide variety of research questions
around the UNSC and speechmaking in the Council (for instance \cite{Sakamoto2023,turco2024speaking,Verbeek2024,Scherzinger2022,Mesquita2024,rodven-eide-etal-2023-unsc,Jankin2024,2023jankaus,eckhard2023international}).

Participants to public UNSC meetings are typically representatives of member states, the UN bureaucracy, or other international governmental and non-governmental organizations. 
% Speakers were member state representatives, representatives of the UN bureaucracy, or representatives of other international governmental and non-governmental organizations.

The dataset %It 
complements ongoing efforts to advance text-as-data research on political texts (Grimmer and Stewart 2013) and to systematize text-based and automated analyses of international conflicts \parencite{2003-king-inf_extract}.
Focusing specifically on text and speech-based analyses of the politics and administration of international organizations, this corpus thus complements the speech transcripts of the UN General Debate corpus (1946-2023) \parencite{2017-baturo-ungd_corpus}, the corpus of speeches of the UN Committee on the Peaceful Uses of Outer Space \parencite{2017-pomeroy-un_space} and research that works with speech as the output of international organizations \parencite{2019-squatrito-shaming_by_io}.
%Most recently, this has increased the attention to text and speech-based analyses of the politics and administration of international organizations, from speech transcripts in the UN General Debate corpus \parencite{2017-baturo-ungd_corpus} or the corpus of speeches of the UN Committee on the Peaceful Uses of Outer Space \parencite{2017-pomeroy-un_space} to research that works with speech as the output of international organizations \parencite{2019-squatrito-shaming_by_io}.
Given that the dataset allows distinguishing speech contributions by member states 
from speech contributions by the bureaucracy, the UNSC debate corpus also provides a new source of evidence for the study of the power, authority and influence of international public administrations (IPAs) in global policy-making \parencite{abbott_genschel_snidal_zangl_2015,rulesworld,2017intburea,mgnrsglblchange,2016eckhardinflburea,2018eckhardreform,2006hawkins,Hooghe2015,trondalunpackingio,2021-svanh-intbur,2023-eckhard-performance,2022-knill-ipa,2019-rittb-io,2024-reinalda-handb-io,2022-eckhard-politics}. 
%Abbott et al., 2015; Barnett & Finnemore, 2004; Bauer et al., 2017; Biermann & Siebenhüner, 2009; Eckhard & Ege, 2016; Eckhard et al., 2018; Hawkins et al., 2006; Hooghe & Marks, 2015; Rittberger et al., 2019; Trondal et al., 2010). For the reference list:

The first descriptive statistics included in this paper indicate three broad trends in the UNSC debates since 1992. Firstly,
%The descriptive statistics presented in this paper indicate that three broad trens are at work in USNC debates since 1995: First of all, 
debates in the UNSC are by no means dominated by the exclusive circle of the Permanent Five (the UNSC veto powers). %UNSC veto powers.
Instead, the largest share of speech contributions comes from non-permanent UNSC member states and %as well as 
invited speakers. Secondly, UNSC meetings are becoming increasingly lengthy and complex. %A second observation is the increasing length and complexity of UNSC meetings. 
Beginning in the 2010s, we see a growing number of very long \enquote{open debate} meetings with an extraordinary high number of speakers. During an open debate all interested states are invited to participate and contribute. Topics discussed are usually of general importance such as \enquote{Women, Peace and Security} or the \enquote{Impacts of Climate-related Disasters}. This trend might reflect the growing concern about such globalized policy challenges. A third and final observation regards
% A final observation is about 
the role of the UN administration, i.e., the Secretary General and her staff. Between 1992 and 2023, the UN administration was the sixth most frequent speaker in the UNSC, with a share of speech contribution similar to each of the P5 members.
%very similar share of speech contributions compared to each P5 member. 
If speech would equalize influence, one could even speak of the P6 UNSC members, the UN administration being one of them \parencite{eckhard2023international}.

In the following sections, we %will 
first discuss how we constructed the UN Security Council Debates corpus through automated and semi-automated text extraction (Section \ref{sec:construc}). After that we present the dataset through descriptive statistics (Section \ref{sec:descrip}) and we show how the descriptive statistics reveal both typical UNSC meetings as well as emerging patterns since 2014 (Section \ref{sec:patterns}.
%  before presenting the dataset through descriptive statistics (Section \ref{sec:descrip}). In Section \ref{sec:patterns}, we show how the descriptive statistics reveal both typical UNSC meetings as well as emerging patterns since 2014. 
We conclude with suggestions for future analyses based on the corpus.

\section{Constructing the UN Security Council Debates corpus}\label{sec:construc}
In order to compile a corpus of single speeches from UNSC meeting protocols several processing steps were necessary. All these steps were conducted on publicly available PDF-documents which we downloaded from the website of the UN\footnote{\url{http://research.un.org/en/docs/sc/quick/meetings/2019} (last accessed May 2019)}. Fortunately, these PDF-documents are well-structured and formatted consistently throughout the years allowing us to automate extraction of speeches as well as basic metadata to a great extent.
Between 1992 and \coveredperiodend{} there were over 500 official communiqués listed by the UN. These communiqu\'{e}s were left out of the analysis.
% Within the timespan between 1995 and 2017 question there were 495 official communiques listed by the UN that were left out of further analysis. 
Preprocessing and compilation of the corpus consisted of the following steps which are described in depth in the following: 

\begin{enumerate}
  \item{Extracting raw text from PDF-documents;}
  \item{Cleaning up raw text;}
  \item{Splitting up raw text into distinct speeches;}
  \item{Labeling speeches by speaker’s names and countries (or affiliations).}
\end{enumerate}

\paragraph{Extracting raw text from PDF-documents:} Conversion of PDF-documents into raw text required a conversion of the two-column layout used for the protocols into a one-column representation. This conversion was done with a tool called \textit{k2pdfopt} that is specifically developed to support re-flow of multi-column PDF texts.\footnote{\url{http://www.willus.com/k2pdfopt/} (last accessed May 2019)} From input PDF-documents it produces an image-file of high resolution with a continuous paragraph of text with all white borders removed. To facilitate the next steps, \textit{k2pdfopt} also allows for cropping page headers and footers that contain identifiers of meetings or page-numbers. These image-files were then fed into a tool for optical character recognition (OCR) in order to convert them into text documents. The OCR-tool of choice is called \textit{tesseract}, an open-source tool representing state-of-the-art in translating image files to text documents.\footnote{\url{https://github.com/tesseract-ocr/tesseract} (last accessed May 2019)} It benefits from dictionary-like language models that are used to increase translation performance \parencite{2007_smith_tesseract,2009_smith_tesseract}. 

For recent documents that contain digital text rather than images, converting PDFs to raw text was more straightforward using PDF editors to export them to the DOCX format. DOCX is particularly useful as it effectively detects headers and footers — elements that need to be removed from the transcript — while also preserving the flow of speeches despite the two-column layout. Furthermore, since OCR is not required, this approach allows for faster text import into programming languages such as R.

\paragraph{Cleaning up raw text:} The resulting text documents included some minor errors %had some minor issues 
that had to be corrected. One example %such thing 
were ligatures, long dashes and special space characters. These are non-ASCII characters might hinder further processing since they cause two character strings to look different in machine-consumed Byte-code representation when in fact both character strings contain the exact same words.
%, or special space characters, all of which are non-ASCII characters that might hinder further processing: These special characters cause two character strings to look differently in machine-consumed Byte-code representation when in fact both character strings contain the exact same words. 
Hence, such special characters were removed using pattern matching.

\paragraph{Splitting up raw text into distinct speeches:} Since UNSC protocols are consistently well-structured they can be divided into distinct speeches using certain recurring patterns of text that mark the beginning of a speech. This pattern describes a form of address or the noun-marker \enquote{The} at the beginning of a line
%This pattern describes at a beginning of a line a form of address or the noun-marker \enquote{The} 
followed by at least one word up to a colon. Optionally, before the colon, there are one or two insertions in parentheses containing a country and/or a hint in which language the speech was held. A few examples illustrate this pattern: 
\begin{itemize}
  \item{\enquote{The President:}}
  \item{\enquote{The Secretary-General}}
  \item{\enquote{Mr. Levitte (France) (spoke in French):}}
  \item{\enquote{Ms. Schoulgin Nyoni (Sweden):}}
\end{itemize}

The adresses were used to identify the beginning of a speech:
\begin{table}[h!]
\centering
\begin{tabular}{|l|l|l|l|l|}
\hline
The President         & Baron      & Monsignor & Prince       & Dr. \\
The Secretary-General & Baroness   & Sir       & Crown Prince &     \\
Lieutnant             & Nana       & Miss      & Princess     &     \\
Major                 & Dato       & Mr.       & Lord         &     \\
General               & Datuk      & Ms.       & Sheikh       &     \\
Judge                 & Archbishop & Mrs.      & King         &     \\ \hline
\end{tabular}%
\caption{Words that mark beginning of new speeches}
\label{tab:begspch}
\end{table}
%~ \begin{multicols}{4}
%~ \begin{itemize}
  %~ \item{The President}
  %~ \item{The Secretary-General}
  %~ \item{Lieutnant}
  %~ \item{Major}
  %~ \item{General}
  %~ \item{Judge}
  %~ \item{Prince}
  %~ \item{Crown Prince}
  %~ \item{Princess}
  %~ \item{Lord}
  %~ \item{Sheikh}
  %~ \item{King}
  %~ \item{Baron}
  %~ \item{Baroness}
  %~ \item{Nana}
  %~ \item{Dato}
  %~ \item{Datuk}
  %~ \item{Archbishop}
  %~ \item{Monsignor}
  %~ \item{Sir}
  %~ \item{Dr.}
  %~ \item{Miss}
  %~ \item{Mr.}
  %~ \item{Ms.}
  %~ \item{Mrs.}
%~ \end{itemize}
%~ \end{multicols}

Sometimes the insertions in parentheses spanned across two lines. Therefore, we first removed line-breaks from lines that contained an opening but no closing parenthesis. Also, line-breaks before a parenthetical insertions containing \enquote{spoke in \dots} were removed. Both steps helped the above-mentioned pattern matching to focus on single lines. 

The matching itself was conducted using regular expressions. For every match a text-document was created containing all text up to the next match or the end of the document. The text-document was given a sequential number to keep the order of the speeches. 

Sometimes, speeches were interrupted by a vote initiated by the speaker. In these cases the protocols would list the corresponding results and the speech continued afterwards.
%The protocols of course listed the corresponding results and the speech continued afterwards. 
To prevent such speeches from being split across two documents, two consecutive speech-documents containing a speech from the same speaker were combined. This resulted in \numspeeches{} speech documents. 

\paragraph{Labeling speeches by speakers' names and countries or affiliations:} In order to be able to annotate speeches with countries or affiliations two approaches were necessary. First, for regular attendees countries had to be extracted from the first page of every protocol that contained the official list of participants. Second, for external speakers invited by the UNSC presidency during a session indications in parentheses in the text had to be evaluated. 

From the first page of every protocol the list of attendees was extracted using pattern matching. To be able to do this the first page was separated from the rest of the protocol before applying the OCR procedure, mainly because the first page required slightly different parameter settings. After applying a tailored OCR to the first page, the speaker-to-country-mapping was extracted. This resulted in a dictionary associating speakers to the nation they represent. As a result, every speech in the corpus could be associated with a country name. This was especially useful for cases in which the country was not indicated after the speaker's name at the beginning of a speech. 

Labeling attendees who were invited during the meeting and did not appear in the official list of participants on the first page was less straightforward. In %However, in 
some debates the number of this type of attendees even exceeded %s 
the number of regular UNSC members. These other attendees were %are 
invited by the UNSC President chairing a meeting
%by inviting them 
to \enquote{take a seat at the table}. For some of these attendees a country or affiliation was given in parentheses at the beginning of the speech that could be automatically extracted. For other invited speakers we had to enter their affiliation manually.

\section{Descriptive Statistics of the UN Security Council Debates Corpus}\label{sec:descrip}
\begin{figure}[ht]
\centering
\begin{subfigure}{.49\textwidth}
  \centering
	  \includegraphics[width=\textwidth]{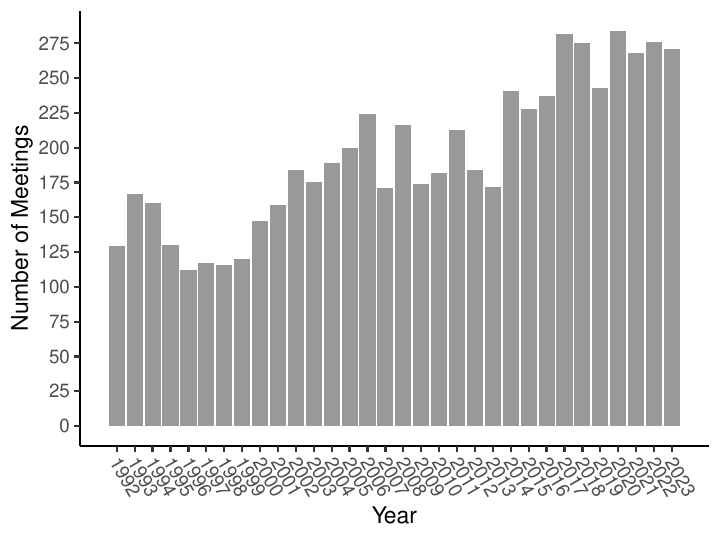}
  \caption{Numbers of meetings per year}
  \label{fig:meetperyear}
\end{subfigure}
\begin{subfigure}{.49\textwidth}
  \centering
  \includegraphics[width=\textwidth]{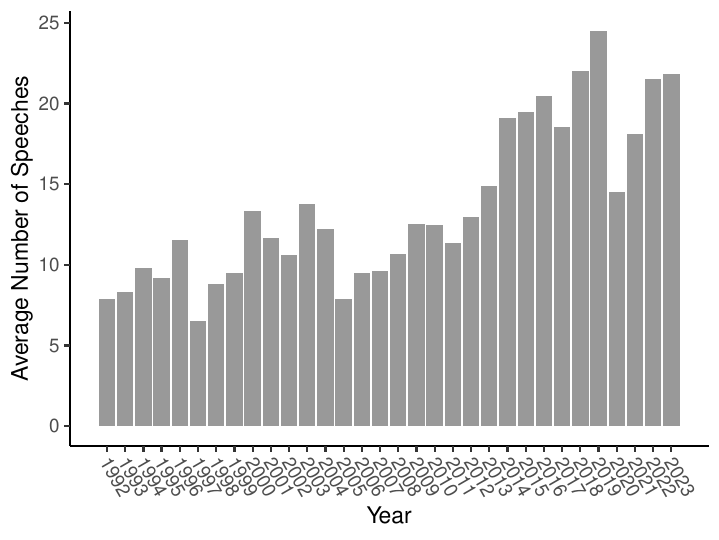}
  \caption{Average number of speeches per meeting}
  \label{fig:avglengthmeetperyear}
\end{subfigure}
\caption{Meetings over time.}
\label{fig:meetings}
\end{figure}

The UNSC corpus consists of \numspeeches{} speeches that were held between January 6, 1992, and December 30, \coveredperiodend. This timeframe covers \nummeetings{} meetings. Figure \ref{fig:meetperyear} shows how the meetings distribute over the years. It can be seen that the number of meetings per year increased over time -- in 1992 our dataset contains 130 meetings whereas for \coveredperiodend{} we count \nummeetingslastyear{} meetings -- but there is variation from year to year. At the same time, the average length of the meetings -- represented by the average number of speeches per meeting -- increased as well from 1992 until the 2020s. A first peak is visible around the Iraq war in 2003 and then there is a steady increase from the mid-2000s until the 2020s,  with a notable gap in the Covid-19 years 2020. Since 2020, both the number of meetings and the length of meetings seem to have reached a plateau.
%, with a first peak around the Iraq war in 2003 and then a steady increase from the mid-2000s until the mid-2010s 
(see Figure \ref{fig:avglengthmeetperyear}). 

\begin{figure}[h!]
\centering
\begin{subfigure}{.49\textwidth}
  \centering
  \includegraphics[width=\textwidth]{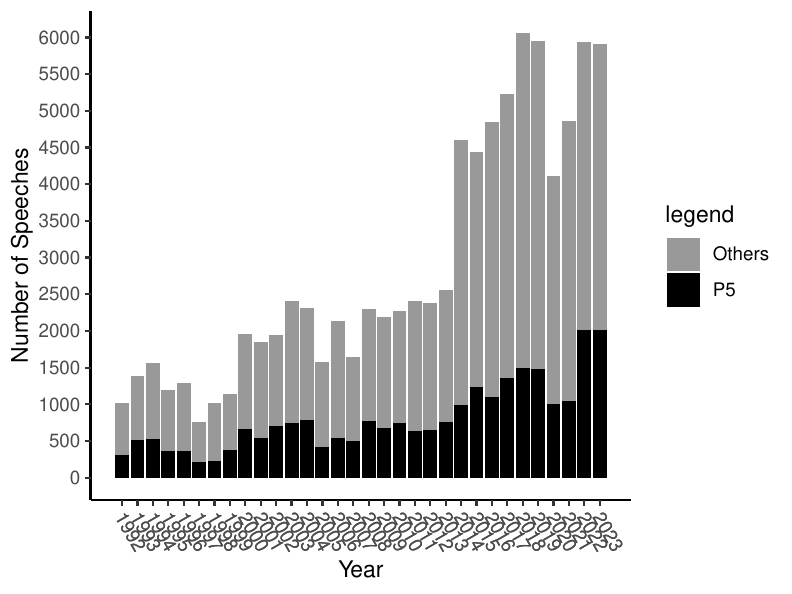}
  \caption{Numbers of speeches per year.}
  \label{fig:nspeeches}
\end{subfigure}
\begin{subfigure}{.49\textwidth}
  \centering
  \includegraphics[width=\textwidth]{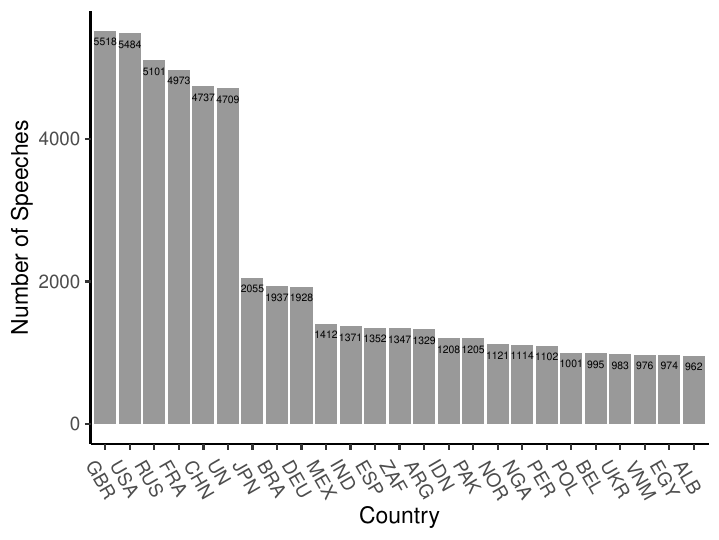}
  \caption{The 25 most frequent speaker affiliations}
  \label{fig:topcountries}
\end{subfigure}
\caption{Numbers of speeches per year.}
\label{fig:speechesovertime}
\end{figure}

The absolute number of UNSC speeches per year is depicted in Figure \ref{fig:nspeeches}. One can identify five major periods: The years from 1992 to 1999; the period from 2000 until 2004;
% four major periods: the period from 2000 until 2004; 
the years 2005 until 2007; the period from 2008 until 2013; and the period starting in 2014. With a view to the number of meetings per year (see Figure \ref{fig:meetings}), one can see that the UN Security Council met less frequently during the first period than during most of the other years, and the total number of speeches reflects this with only about 1,400 speeches per year.
% during the first period the UN Security Council simply met less frequently than during most of the other years, so the total number of speeches  reflects this with only about 1,400 speeches per year.  
Beginning in 2000, a few years into the tenure of UN Secretary-General Kofi Annan, the number of meetings as well as the length of the meetings increased. Most notably, this period includes the wars in Afghanistan and Iraq.
%This period includes Afghanistan war and the Iraq are, most notably. 
For the five years from 2000 up to and including 2004, our dataset lists more than 2500 speeches annually. After 2004, the length of meetings drops briefly (Figure \ref{fig:nspeeches}) and so does the number of speeches. From 2008 onwards, with the average length of meeting increasing again, the number of speeches is stable around 3,000 per year. After 2013, the total number of speeches increases significantly, reaching 4,757 in 2014 and \numspeecheslastyear{} by \coveredperiodend{}, with a drop during the Covid-19 years just before. 
% not sure we can draw this as a causal conclusion: `This is mainly driven by the average length of meetings'. Is the length of a meeting determined by the number of speech or the number of speeches by the length of the meeting? Or both by the agenda and invitations?}
This could reflect an increasing importance of the UNSC over the past years as well as a higher level of conflict so that there is more need for debate. During this period, we also see an increasing number of open debates, many with a very high number of invited speakers who represent states who were not a member of the UNSC at the time or from other international governmental and non-governmental organizations. Thus, the nature of UNSC meetings appears to have shifted %rather 
fundamentally since 2014.

Using the metadata that allows us to identify the country or affiliation of each speaker, Charts \subref{fig:nspeeches}) and \subref{fig:topcountries}) in Figure \ref{fig:speechesovertime}
%\ref{fig:tokensovertime} 
provide insights about the distribution of speakers over time and across countries. In Figure \ref{fig:nspeeches}, the number of speeches of the five permanent members of the UNSC (the P5) is highlighted in black. The relative share of P5 speeches broadly follows the total number of speeches in each year until 2013. From 2014 onwards, however, the increase of the total number of speeches is not followed with a similar increase in the share of P5-speeches.
% Then, the increase of the total number of speeches from 2014 onwards is not followed with a similar increase in the share of P5-speeches. 
As indicated above, this reflects an increased activity of the elected 10 (E10) UNSC members and the increasing number of open debates in which non-UNSC members are invited to speak.

Overall, the P5 are also the top 5 speakers during the 1995-\coveredperiodend{} period as can be seen in  %the
Chart \subref{fig:topcountries}) of Figure \ref{fig:speechesovertime}. While this is not surprising given the permanent presence %in 
of the P5 in the UNSC, it is noticeable that speakers representing the United Nations bureaucracy -- the Secretaries-General, their Special Representatives or other high-level officials from other UN agencies and bodies -- speak almost as often as the P5 members and much more frequently than even important non-permanent members of the UNSC such as Japan or Germany.\footnote{For a complete list of contributors, please refer to Table \ref{tab:countries} at the end of this paper.} This underlines that the UN bureaucracy could have a significant influence on shaping UNSC debates. We know however that the degree to which UN Secretaries-General have made use of this potential to influence the debates varies significantly over time
% The degree to which, for example, UN Secretaries-General have made use of this influence potential varies significantly over time 
(see the contributions to a recent edited volume on the UNSC-SG relationship  \parencite{2018-froehlich-williams}). The present dataset can provide insights into how this influence potential extends beyond the Secretaries-General.

\begin{figure}[h!]
    \centering
    \includegraphics[width=\linewidth]{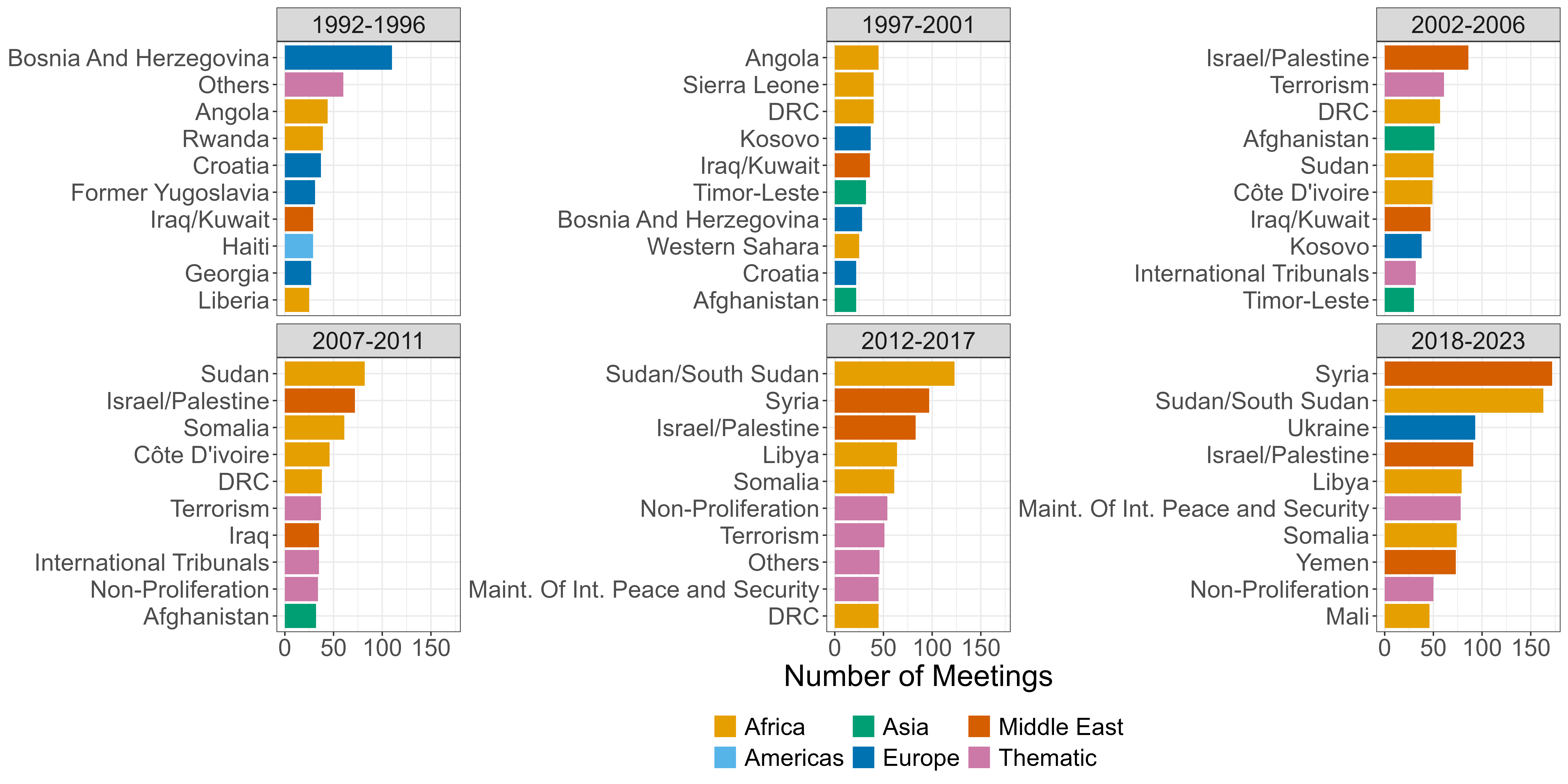}
    \caption{Number of meetings per agenda item and time period.}
    \label{fig:agenda_item}
\end{figure}

Next, Figure \ref{fig:agenda_item} illustrates the evolving focus of the United Nations Security Council (UNSC) over time, highlighting the number of meetings dedicated to specific agenda items across different periods from 1992 to 2023. The agenda item labels shown in the graph are standardized and summarized versions of the official agenda items, based on the Repertoire of the Practice of the Security Council. For instance, items such as "Report of the Secretary-General in Sudan", "Sudan", and "Report of the Secretary-General on the Sudan and South Sudan" have been consolidated under the label "Sudan/South Sudan", following the terminology used in the Repertoire, which categorizes them as "Reports of the Secretary-General on the Sudan and South Sudan". The label "Others" encompasses meetings on various topics, including the admission of new members, commemorative events, ICJ elections, and similar matters.

The early 1990s were dominated by discussions on Bosnia and Herzegovina, reflecting the intense conflicts in the Balkans during that time. As the years progressed, African issues such as those related to Sudan, the Democratic Republic of Congo (DRC), and Somalia became increasingly prominent, particularly from 1997 to 2011. Thematic issues like terrorism and non-proliferation emerged as significant topics in the early 2000s and have maintained consistent attention in subsequent years. Post-2018, the agenda reflects ongoing global concerns, with Ukraine appearing as a key focus alongside continued discussions on Syria, Sudan, and thematic issues such as the "Maintenance of International Peace and Security." This progression underscores the dynamic nature of the UNSC's priorities, shaped by shifting geopolitical crises and global security challenges.

\begin{figure}[h!]
\centering
\begin{subfigure}{.49\textwidth}
  \centering
  \includegraphics[width=\textwidth]{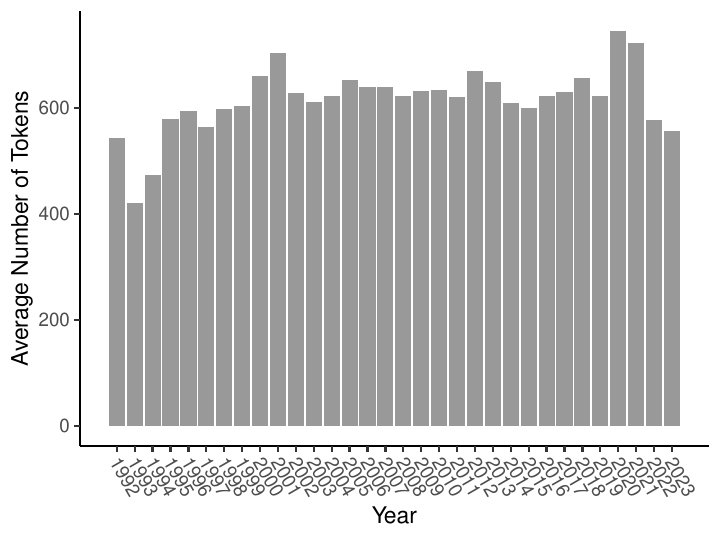}
  \caption{Average number of tokens per speech per year.}
  \label{fig:ntokensperyear}
\end{subfigure}
\begin{subfigure}{.49\textwidth}
  \centering
  \includegraphics[width=\linewidth]{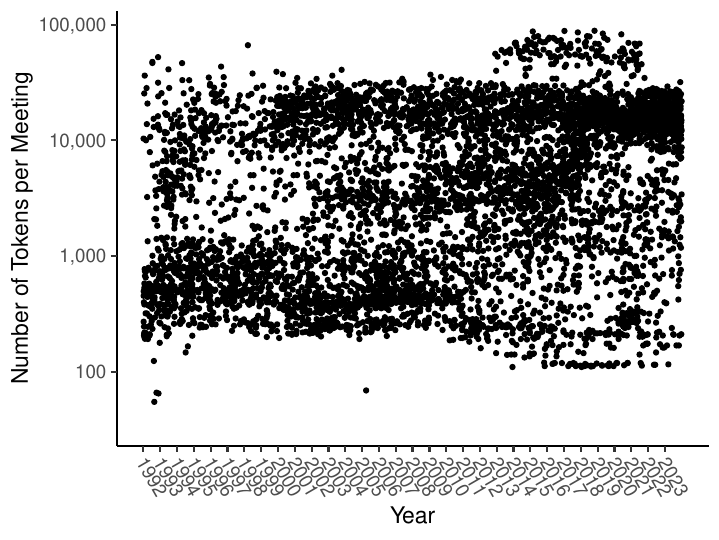}
  \caption{Tokens per meeting -- logarithmic scale.}
  \label{fig:ntokenssingle}
\end{subfigure}
\caption{Numbers of tokens over time.}
\label{fig:tokensovertime}
\end{figure}

In Figure \ref{fig:tokensovertime}, we show the evolution of the lengths of meetings in terms of the number of tokens
of speeches. Tokens allow analysis and comparisons of the length of each speech in more detail. To obtain their number a standard tokenization procedure is applied which means sentences are split at punctuation marks as well as at whitespaces. We rely on the excellent \textit{R}-library quanteda to apply this technique \parencite{quanteda}.

In Figure \ref{fig:ntokensperyear} the numbers of tokens are aggregated over the years and the mean is calculated for every year separately. It can be seen that the length of speeches remains quite stable over time. However, when representing the number of tokens per meeting in Figure \ref{fig:ntokenssingle}, where each dot represents the length of a single meeting measured in the number of tokens, there are two trends observable: First, the total number of tokens is increasing, which mirrors observations on the increasing number of meetings and speeches per meeting summarized above. This is also reflected in the detailed aggregation of the numbers of tokens per year in Table \ref{tab:tokens}. 

\input{num_tokens_per_year.tex}

Second, the diversity of UNSC meeting types increases over time: in the second half of the 1990s, there are either very short (with 1,000 or less tokens) or longer meetings (with around 10,000 tokens); from the mid-2000s onward a medium-size type of meeting (with around 5,000 tokens) emerges between the long and very short meetings; and starting in the mid-2010s the length of meetings starts to spread considerably, from extremely short to a new type of very long meeting with around 80,000 tokens. The latter confirms the observation of the increasing use of open debates in the UNSC with a large number of speakers.

\section{Meeting Patterns of the UNSC: From Formalized Discussions to Open Debates?}\label{sec:patterns}
The descriptive statistics outlined above reveal interesting meeting patterns that we can see from the UN Security Council Debate corpus.

\begin{figure}[h!]
\includegraphics[width=\textwidth]{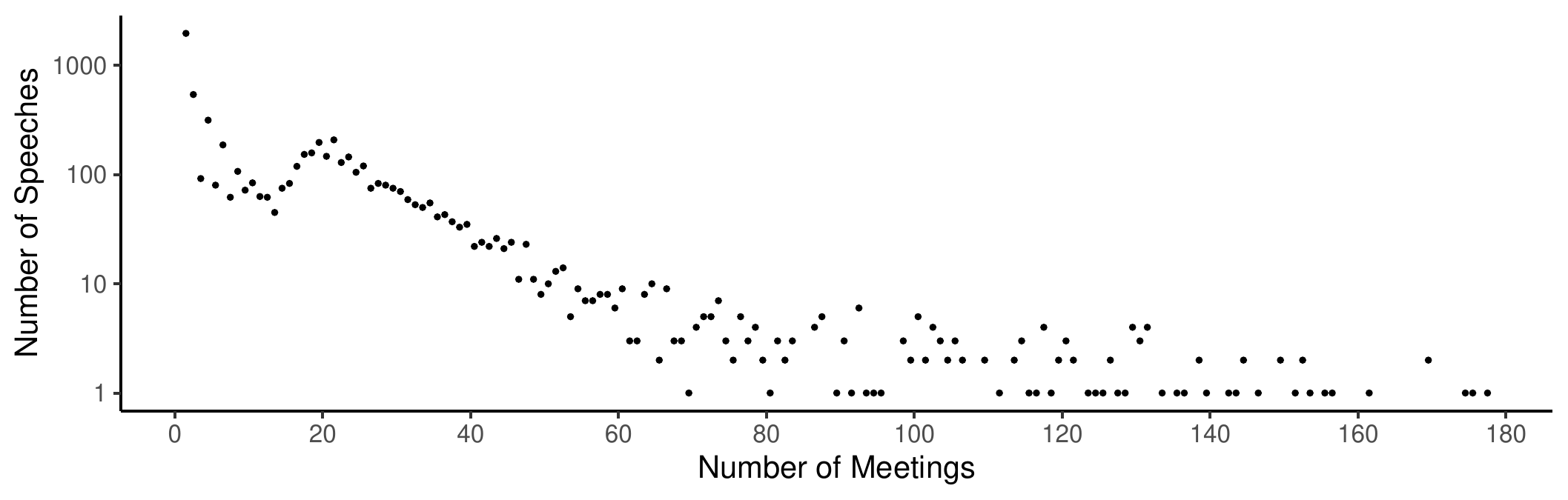}
  \caption{Numbers of speeches per meeting -- dots indicate the number of meetings with a certain number of speeches. Logarithmic scaling on the y-axis.}
  \label{fig:spchpermeetinghist}
\end{figure}

\begin{figure}[h!]
	\includegraphics[width=\textwidth]{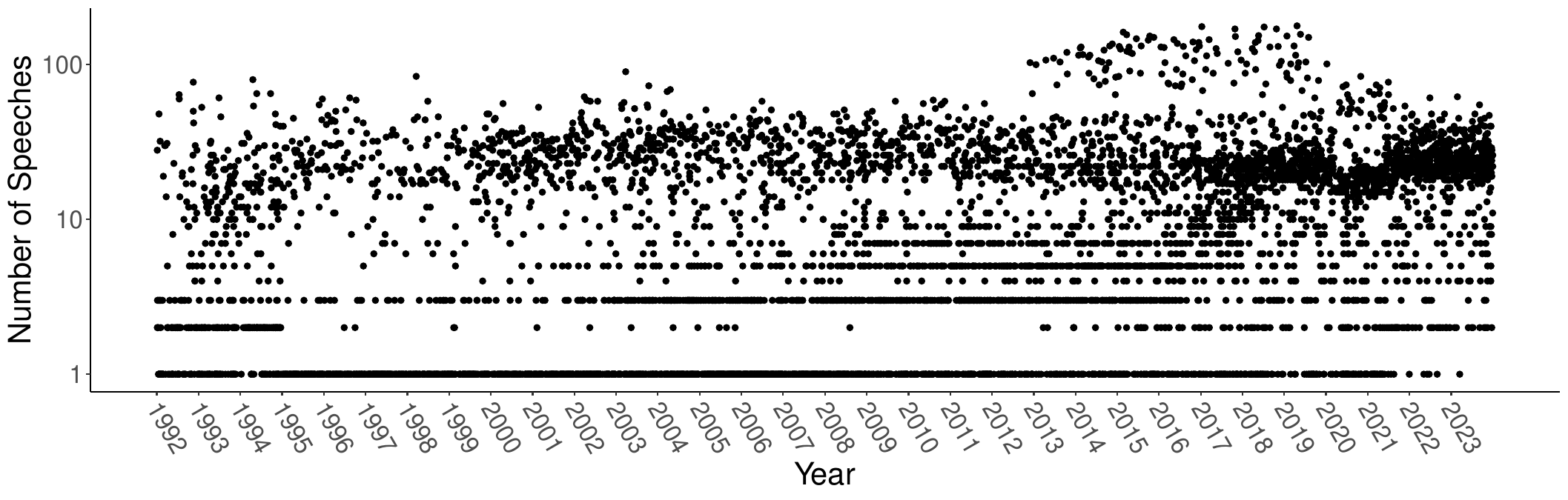}
	\caption{Numbers of speeches over time -- dots indicate the number of speeches held in a meeting}
	\label{fig:speechesovertimeloghist}
\end{figure}

They show (see Figure \ref{fig:spchpermeetinghist}) that there are, broadly speaking, four types of meetings, with the first two being by far the most frequent:

\begin{itemize}
  \item{the \textbf{adoption meeting} with a single speech. In this meeting type
 %, in which 
the Presidency simply calls for a vote on a resolution negotiated behind closed doors without any further discussion; }
  \item{the \textbf{information meeting} with 2-3 speeches. In this meeting type
%, in which  
  the Presidency invites a member or other speakers, such as a representative from the UN bureaucracy, to present a point of view, an explanation of vote, or a report but without any discussion;}
  \item{the \textbf{limited formal debate} with 15 to 40 speeches which peak in frequency around 20 speeches per session.
  % , i.e. meetings with 15 to 40 speeches, which peak in frequency around 20 speeches per session. 
  These are meetings in which some, or all, members and a limited number of UN bureaucracy speakers or other guests each hold one substantive speech, each introduced by a short formal intervention of the presidency inviting the next speaker;}
  \item{the \textbf{intensive or open debate} which extends
  %  , which extend 
  from shorter meetings with 40-60 speeches to very long debates with 100-180 speeches. hese meetings, excluding the usual, formal interventions by the presidency between each speech, include about 50-90 substantive interventions.
  %   Including the formal interventions by the presidency, usually between each speech, this means there are about 50-90 substantive interventions in these longer sessions. 
  This also includes
  %   can also include 
  meetings with more intensive back-and-forth between members on controversial topics such as the Iraq war and the overall Situation in the Middle East,
% (e.g. the Iraq war, the Situation in the Middle East, etc.) (if change not accepted, note that etc. after an i.e. does not make sense)
 and open debates in which all UN members as well as invited non-members (e.g. non-governmental actors) can speak.}
\end{itemize}
As discussed in the previous section, the diversity of meeting patterns also increases over time. Questions that remain open for now are
%One open question is
when and why these patterns emerge and how they change the UNSC debates. For example, with the number of intensive and open debate meetings increasing visibly between 2013 and 2020 -- with all of the sessions with 100 speeches and more taking place in these years (see Figure \ref{fig:speechesovertimeloghist}) -- the dataset in this period includes a much higher diversity of speakers. After 2020, the length of meetings falls back to the pre 2013 pattern, with no extremely long meetings but a high density of intermediate-length meetings.

These patterns make the present dataset interesting for research beyond the P5, and even beyond the non-permanent members, as it may reveal how new arguments, topics and debate styles emerge and disappear and how they relate to broader geopolitical dynamics.

%This makes the present dataset interesting for research beyond the P5 and even beyond the non-permanent members, as it may reveal how new arguments, topics, and debate styles emerge. Potentially, this is not only visible in the open debates but also in the more traditional formats of UNSC meetings. In any case, this trend towards longer open debates could indicate the urge of the UNSC, or at least of some of its members, to open up, not just towards the public but also towards the broader membership of the United Nations that has called to reform the UNSC, which in essence still reflects the geopolitical constellation of the mid-1940s.

\section{Conclusions and outlook}
The UN Security Council Debates corpus presented in this paper offers a further
%is a new 
step towards a quantified understanding of speech and language used in international organizations in general and in the United Nations in particular. It invites research that uses %both
 the entire corpus or %but also 
 relevant subsets of the corpus, such as the \enquote{The UN Security Council debates on Afghanistan corpus} \parencite{2018_unscafg} that we have analyzed elsewhere \parencite{2018_unsc_discurs}. Beyond a focus on UNSC member states and non-governmental organisations, the corpus should also be of interests
 % Analysis of the different roles of member states and international public administrations in these debates also speaks 
  to research on the policy-making influence of international bureaucracies %\parencite{abbott_genschel_snidal_zangl_2015,rulesworld,2017intburea,mgnrsglblchange,2016eckhardinflburea,2018eckhardreform,2006hawkins,Hooghe2015,Rittberger2003,trondalunpackingio}
\parencite{abbott_genschel_snidal_zangl_2015,rulesworld,2017intburea,mgnrsglblchange,2016eckhardinflburea,2018eckhardreform,2006hawkins,Hooghe2015,trondalunpackingio,2021-svanh-intbur,2023-eckhard-performance,2022-knill-ipa,2019-rittb-io,2024-reinalda-handb-io,2022-eckhard-politics,eckhard2023international,2023jankaus}.

As noted before, the \coveredperiodend{} UNSC corpus complements and extends the
 established UN-related speech corpora, such as the UN General Debate corpus \parencite{2017-baturo-ungd_corpus} yet offers interesting new possibilities and avenues for research. % :  
First, the UNSC debates represent the everyday high-level diplomatic politics of the UN, whereas the UN General Debates takes place only once a year and rather reflect %s
 national leaders' broad visions of global politics. Second, the UNSC Debates corpus %also
  allows to focus on specific topics of international security where all speakers relate to this topic but may show different levels of interest, emotion or sub-topic focus, such as the topics 
   \enquote{Women, Peace and Security} and the \enquote{Impacts of Climate-related Disasters}.
      % sentence order + grammer changed, original:  Second, the UNSC Debates corpus also allows to focus on specific topics of international security, such as meetings and open debates on “Women, Peace and Security” or on the “Impacts of Climate-related Disasters”, where all speakers relate to this topic but may show different levels of interest, emotion or sub-topic focus.
   We expect that this will allow for %a 
   more fine-grained analyses 
    of the trends in international politics as well as better research into the dynamics of state coalitions over 
    time and across different topics.%thematic subjects.

Overall, further analyses of the UNSC corpus making use of the full range of natural language processing and quantitative text analysis methods can provide a new understanding of the history and dynamics of international politics as reflected in the debates of the UNSC. This corpus also allows for specific types of analyses, given that actors' speeches in this body
%This view will certainly be specific, given that the speeches actors make in this body
 constitute carefully crafted foreign policy statements. % that they want to communicate to other actors and the general public. 
Rather than a downside, % But 
this makes analyses %is 
of UNSC debates particularly interesting. Future research may answer a range of unexplored questions, such as: Who % including questions such as who
 speaks about which (common) topics;
  which topics and meetings arouse emotions more than others; %are  more emotional,
which actors are more diplomatic;
 who is influential in framing debates;
  to what extent do speech contributions reflect voting patterns; and, finally, which issues and trends extend beyond specific meeting topics and spread across all or even beyond UNSC debates? In short, we are convinced this new corpus offers great potential for further research into international security in general and the role of international organisations and the UNSC in specific.

%The UN Security Council Debates corpus presented in this paper is a new step towards a quantified understanding of speech and language used in international organizations in general and in the United Nations in particular. It invites research that uses both the entire corpus but also relevant subsets, such as the \enquote{The UN Security Council debates on Afghanistan corpus} \parencite{2018_unscafg} that we have analyzed elsewhere \parencite{2018_unsc_discurs}. It adds to already established UN-related speech corpora, such as the UN General Debate corpus \parencite{2017-baturo-ungd_corpus}: First, the UNSC debates represent the everyday high-level diplomatic politics of the UN, whereas the UN General Debate only takes place only once a year and rather reflects national leaders broad visions of global politics. The UNSC Debates corpus also allows to focus specific topics, such as meetings and open debates on Women, Peace and Security (WPS), where all speakers relate to this topic but may show different levels of interest, emotion or sub-topic focus. We expect that this will allow for a more fine-grained analysis of state coalitions across time and across different meeting topics. Further analyses making use of the full range of natural language processing and quantitative text analysis methods can reveal which topics and meetings are more emotional, which countries are more diplomatic, which actors are more influential in framing debates, or which issues and trends extend beyond specific meeting topics and spread across all UNSC debates.

\vspace{15pt}
\printbibliography

\pagebreak
\section*{Appendix}
\input{countries_and_speeches_totale.tex}

\end{document}

%% file: num_tokens_per_year.tex
% latex table generated in R 4.4.2 by xtable 1.8-4 package
% Thu Dec 19 13:58:24 2024
\begin{table}[ht]
\centering
\begin{tabular}{||r|r||r|r||r|r||r|r||}
  \hline
Year & Tokens & Year & Tokens & Year & Tokens & Year & Tokens \\ 
  \hline
1992 & 551916 & 2000 & 1239341 & 2008 & 1409027 & 2016 & 3024848 \\ 
  1993 & 585099 & 2001 & 1210568 & 2009 & 1321240 & 2017 & 3301461 \\ 
  1994 & 730739 & 2002 & 1184092 & 2010 & 1400897 & 2018 & 3985512 \\ 
  1995 & 649778 & 2003 & 1453367 & 2011 & 1475236 & 2019 & 3739808 \\ 
  1996 & 774510 & 2004 & 1377188 & 2012 & 1570675 & 2020 & 3157480 \\ 
  1997 & 404183 & 2005 & 1007380 & 2013 & 1635915 & 2021 & 3615885 \\ 
  1998 & 575870 & 2006 & 1325446 & 2014 & 2794699 & 2022 & 3698668 \\ 
  1999 & 643819 & 2007 & 1022082 & 2015 & 2665481 & 2023 & 3621860 \\ 
   \hline
\end{tabular}
\caption{Number of Tokens per Year} 
\label{tab:tokens}
\end{table}

%% file: countries_and_speeches_totale.tex
\begingroup\fontsize{7}{9}\selectfont

\begin{longtable}[t]{>{\raggedright\arraybackslash}p{8.5em}l>{\raggedright\arraybackslash}p{8.5em}l>{\raggedright\arraybackslash}p{8.5em}l>{\raggedright\arraybackslash}p{8.5em}l}
\caption{\label{tab:countries}Countries and the number of speeches}\\
\toprule
Country & Count & Country & Count & Country & Count & Country & Count\\
\midrule
\endfirsthead
\caption[]{Countries and the number of speeches \textit{(continued)}}\\
\toprule
Country & Count & Country & Count & Country & Count & Country & Count\\
\midrule
\endhead

\endfoot
\bottomrule
\endlastfoot
\cellcolor{gray!10}{Abductees' Mothers Association} & \cellcolor{gray!10}{1} & \cellcolor{gray!10}{ACADESAN} & \cellcolor{gray!10}{1} & \cellcolor{gray!10}{ACCORD} & \cellcolor{gray!10}{2} & \cellcolor{gray!10}{Action Against Hunger} & \cellcolor{gray!10}{1}\\
ADVOX & 1 & Afghan High Peace Council & 1 & Afghan Independent Human Rights Commission & 1 & Afghan Institute For Strategic Studies & 1\\
\cellcolor{gray!10}{Afghan Women Skills Development Centre} & \cellcolor{gray!10}{1} & \cellcolor{gray!10}{Afghan Women's Network} & \cellcolor{gray!10}{2} & \cellcolor{gray!10}{Afghan Youth} & \cellcolor{gray!10}{2} & \cellcolor{gray!10}{Afghanistan} & \cellcolor{gray!10}{191}\\
Afghanistan Independent Human Rights Commission & 3 & Afghanistan Research And Evaluation Unit & 1 & Afghans For Progressive Thinking & 1 & Afia Mama (Organization : Democratic Republic of Congo) & 1\\
\cellcolor{gray!10}{African Development Bank} & \cellcolor{gray!10}{1} & \cellcolor{gray!10}{African National Congress} & \cellcolor{gray!10}{2} & \cellcolor{gray!10}{African Union} & \cellcolor{gray!10}{266} & \cellcolor{gray!10}{African Unity} & \cellcolor{gray!10}{5}\\
\addlinespace
African Women In Dialogue & 1 & African Women Leaders Network & 1 & Al Amal Fund & 1 & Al Azhar Islamic Research Academy & 1\\
\cellcolor{gray!10}{Al-Amal Association} & \cellcolor{gray!10}{1} & \cellcolor{gray!10}{Albania} & \cellcolor{gray!10}{962} & \cellcolor{gray!10}{Algeria} & \cellcolor{gray!10}{326} & \cellcolor{gray!10}{Alliance of Civilizations} & \cellcolor{gray!10}{1}\\
Alliance Of Small Island States & 1 & Amal-Tikva & 1 & Amani Africa & 2 & Andorra & 10\\
\cellcolor{gray!10}{Angaza Institute} & \cellcolor{gray!10}{1} & \cellcolor{gray!10}{Angola} & \cellcolor{gray!10}{590} & \cellcolor{gray!10}{Anthropic (Firm)} & \cellcolor{gray!10}{1} & \cellcolor{gray!10}{Antigua And Barbuda} & \cellcolor{gray!10}{7}\\
Arab American Institute & 1 & Arab Human Rights Foundation & 2 & Arab Maghreb Union & 1 & Argentina & 1329\\
\addlinespace
\cellcolor{gray!10}{Armenia} & \cellcolor{gray!10}{124} & \cellcolor{gray!10}{ASEAN} & \cellcolor{gray!10}{6} & \cellcolor{gray!10}{Asha Gelle Foundation} & \cellcolor{gray!10}{1} & \cellcolor{gray!10}{Assistance Mission For Africa} & \cellcolor{gray!10}{1}\\
Association Alliance For Peace And Security & 1 & Association Des Femmes Peules Autochtones Du Tchad & 1 & Association IMAD pour la jeunesse et la paix & 1 & Association Of African Women For Research And Development & 1\\
\cellcolor{gray!10}{Association Of Central African Women Lawyers} & \cellcolor{gray!10}{1} & \cellcolor{gray!10}{Association Of Indigenous Cabildos Of The North Of Cauca} & \cellcolor{gray!10}{1} & \cellcolor{gray!10}{Association Protection Sahel} & \cellcolor{gray!10}{1} & \cellcolor{gray!10}{Asuda for Combating Violence Against Women (Organization : Iraq)} & \cellcolor{gray!10}{1}\\
Australia & 694 & Austria & 429 & Azerbaijan & 478 & B'tselem & 1\\
\cellcolor{gray!10}{Bahamas} & \cellcolor{gray!10}{6} & \cellcolor{gray!10}{Bahrain} & \cellcolor{gray!10}{216} & \cellcolor{gray!10}{Bangladesh} & \cellcolor{gray!10}{602} & \cellcolor{gray!10}{Barbados} & \cellcolor{gray!10}{8}\\
\addlinespace
Belarus & 59 & Belgium & 995 & Belize & 7 & Benin & 257\\
\cellcolor{gray!10}{Bhutan} & \cellcolor{gray!10}{1} & \cellcolor{gray!10}{Bolivia (Plurinational State Of)} & \cellcolor{gray!10}{751} & \cellcolor{gray!10}{Bosnia And Herzegovina} & \cellcolor{gray!10}{404} & \cellcolor{gray!10}{Botswana} & \cellcolor{gray!10}{266}\\
Brazil & 1937 & Brian Urquhart Center For Peace Operations & 1 & Brunei Darussalam & 11 & Bulgaria & 276\\
\cellcolor{gray!10}{Burkina Faso} & \cellcolor{gray!10}{320} & \cellcolor{gray!10}{Burundi} & \cellcolor{gray!10}{97} & \cellcolor{gray!10}{Butterflies With New Wings} & \cellcolor{gray!10}{1} & \cellcolor{gray!10}{Cabo Verde} & \cellcolor{gray!10}{2}\\
Cambodia & 34 & Cameroon & 334 & Canada & 686 & Cape Verde & 125\\
\addlinespace
\cellcolor{gray!10}{Carabinieri Command For The Protection Of Cultural Heritage} & \cellcolor{gray!10}{1} & \cellcolor{gray!10}{CARE} & \cellcolor{gray!10}{9} & \cellcolor{gray!10}{Caribbean Community} & \cellcolor{gray!10}{2} & \cellcolor{gray!10}{Carnegie Endowment for International Peace (Washington, D.C.)} & \cellcolor{gray!10}{1}\\
Center For Civil Liberties & 1 & Center For Civil Society And Democracy & 1 & Center For Civilians In Conflict & 1 & Center For Geostrategic Studies & 1\\
\cellcolor{gray!10}{Center On International Cooperation} & \cellcolor{gray!10}{1} & \cellcolor{gray!10}{Central African Republic} & \cellcolor{gray!10}{83} & \cellcolor{gray!10}{Centre For Energy And Climate Of The French Institute Of International Relations} & \cellcolor{gray!10}{1} & \cellcolor{gray!10}{Centre For Inclusive Governance, Peace And Justice} & \cellcolor{gray!10}{2}\\
Centre For Strategic Communication And Information Security & 1 & Centre National D’études Stratégiques Et De Sécurité & 1 & Chad & 347 & Challenges Forum & 1\\
\cellcolor{gray!10}{Charmaghz} & \cellcolor{gray!10}{1} & \cellcolor{gray!10}{Chile} & \cellcolor{gray!10}{962} & \cellcolor{gray!10}{China} & \cellcolor{gray!10}{4737} & \cellcolor{gray!10}{Chinese Academy of Sciences} & \cellcolor{gray!10}{1}\\
\addlinespace
Civil Society & 135 & Coalition Des Femmes Leaders Nord, Sud Et Centre Du Mali & 1 & Coalition Des Organisations De La Société Civile D’afrique Centrale Pour La Préservation De La Paix, La Prévention Des Conflits, La Resolution Et La Transformation Des Crises & 1 & Collective Security Treaty Organization & 5\\
\cellcolor{gray!10}{Colombia} & \cellcolor{gray!10}{856} & \cellcolor{gray!10}{Columbia University} & \cellcolor{gray!10}{2} & \cellcolor{gray!10}{Commission Électorale Nationale Indépendante} & \cellcolor{gray!10}{3} & \cellcolor{gray!10}{Commission For The Clarification Of Truth, Coexistence And Non-Repetition Of Colombia} & \cellcolor{gray!10}{1}\\
Commission of the Gulf of Guinea & 1 & Committee On The Exercise Of The Inalienable Rights Of The Palestinian People & 56 & Common Cause Network & 1 & Commonwealth & 2\\
\cellcolor{gray!10}{Commonwealth Of Independent States} & \cellcolor{gray!10}{5} & \cellcolor{gray!10}{Commonwealth Of Nations} & \cellcolor{gray!10}{3} & \cellcolor{gray!10}{Communauté économique des Etats de l'Afrique centrale} & \cellcolor{gray!10}{2} & \cellcolor{gray!10}{Community Development Association Of The Sudan} & \cellcolor{gray!10}{1}\\
Community Empowerment For Progress Organization & 2 & Community Of Portuguese-Speaking Countries & 11 & Community Of Sant'egidio & 2 & Community-Based Initiative & 1\\
\addlinespace
\cellcolor{gray!10}{Comoros} & \cellcolor{gray!10}{7} & \cellcolor{gray!10}{Comprehensive Nuclear-Test-Ban Treaty Organization} & \cellcolor{gray!10}{1} & \cellcolor{gray!10}{Comunità di Sant'Egidio} & \cellcolor{gray!10}{1} & \cellcolor{gray!10}{Concertation des collectifs des associations féminines de la région des Grands-Lacs} & \cellcolor{gray!10}{1}\\
Conference Episcopale Nationale Du Congo & 1 & Conflict Armament Research & 1 & Congo & 242 & Congolese Association For Access To Justice & 1\\
\cellcolor{gray!10}{Conseil National De La Jeunesse Centrafricaine} & \cellcolor{gray!10}{1} & \cellcolor{gray!10}{Consortium of International Agricultural Research} & \cellcolor{gray!10}{1} & \cellcolor{gray!10}{Control Arms Governance Board} & \cellcolor{gray!10}{2} & \cellcolor{gray!10}{Cook Islands} & \cellcolor{gray!10}{1}\\
Coordinating Bureau of the Non-Aligned Countries & 3 & Coordination Centre Of Serbia And Montenegro And Of The Republic Of Serbia For Kosovo And Metohija & 1 & Corporacion De Investigacion Y Accion Social Y Economica & 2 & Costa Rica & 448\\
\cellcolor{gray!10}{Cote D'ivoire} & \cellcolor{gray!10}{603} & \cellcolor{gray!10}{Council Of Europe} & \cellcolor{gray!10}{2} & \cellcolor{gray!10}{Council of the European Union} & \cellcolor{gray!10}{3} & \cellcolor{gray!10}{COVID-19 Vaccine Country Readiness And Delivery} & \cellcolor{gray!10}{2}\\
\addlinespace
Croatia & 461 & CTBTO Youth Group & 1 & Cuba & 233 & Cyprus & 52\\
\cellcolor{gray!10}{Czech Republic} & \cellcolor{gray!10}{197} & \cellcolor{gray!10}{Czechia} & \cellcolor{gray!10}{10} & \cellcolor{gray!10}{Democratic Party Of South Africa} & \cellcolor{gray!10}{1} & \cellcolor{gray!10}{Democratic People's Republic Of Korea} & \cellcolor{gray!10}{32}\\
Democratic Republic Of The Congo & 173 & Denmark & 373 & Deutsche Post DHL Group & 1 & Djibouti & 196\\
\cellcolor{gray!10}{Dominica} & \cellcolor{gray!10}{2} & \cellcolor{gray!10}{Dominican Republic} & \cellcolor{gray!10}{550} & \cellcolor{gray!10}{Dominican Republic/Germany} & \cellcolor{gray!10}{1} & \cellcolor{gray!10}{Dushirehamwe Association} & \cellcolor{gray!10}{1}\\
East African Community & 1 & Economic Community Of Central African States & 13 & Economic Community Of West African States & 30 & Ecopeace Middle East & 4\\
\addlinespace
\cellcolor{gray!10}{Ecuador} & \cellcolor{gray!10}{628} & \cellcolor{gray!10}{Edesia (Organization)} & \cellcolor{gray!10}{1} & \cellcolor{gray!10}{Egypt} & \cellcolor{gray!10}{974} & \cellcolor{gray!10}{El Salvador} & \cellcolor{gray!10}{70}\\
Elman Peace And Human Rights Centre & 3 & Encadrement Des Femmes Indigenes Et Des Menages Vulnerables & 1 & Enda Energie (Organization : Senegal) & 1 & Enough Project & 1\\
\cellcolor{gray!10}{Equatorial Guinea} & \cellcolor{gray!10}{522} & \cellcolor{gray!10}{Eritrea} & \cellcolor{gray!10}{22} & \cellcolor{gray!10}{Estonia} & \cellcolor{gray!10}{591} & \cellcolor{gray!10}{Ethiopia} & \cellcolor{gray!10}{529}\\
European Centre For Conflict Prevention & 1 & European Council & 1 & European Union & 882 & Facilitation For Peace And Development & 1\\
\cellcolor{gray!10}{Families For Freedom} & \cellcolor{gray!10}{2} & \cellcolor{gray!10}{Farida Global Organization} & \cellcolor{gray!10}{1} & \cellcolor{gray!10}{Federated States Of Micronesia} & \cellcolor{gray!10}{4} & \cellcolor{gray!10}{Female Solidarity For Integrated Peace And Development} & \cellcolor{gray!10}{1}\\
\addlinespace
Femmes Africa Solidarite & 1 & Fezzan Libya Organization & 1 & Fiji & 76 & Financial Action Task Force & 2\\
\cellcolor{gray!10}{Finland} & \cellcolor{gray!10}{64} & \cellcolor{gray!10}{Flyktningerad (Norway)} & \cellcolor{gray!10}{1} & \cellcolor{gray!10}{Fondasyon Je Klere} & \cellcolor{gray!10}{2} & \cellcolor{gray!10}{Food For Humanity} & \cellcolor{gray!10}{1}\\
Former Yugoslav Republic Of Macedonia & 23 & Foundation For Human Rights In South Africa & 1 & Founder And Chairwoman Of The Somali Gender Equity Movement & 1 & France & 4973\\
\cellcolor{gray!10}{Freedom Fund} & \cellcolor{gray!10}{1} & \cellcolor{gray!10}{Freedom Now (United States)} & \cellcolor{gray!10}{1} & \cellcolor{gray!10}{French Red Cross} & \cellcolor{gray!10}{1} & \cellcolor{gray!10}{Fuerzas Armadas Revolucionarias de Colombia} & \cellcolor{gray!10}{1}\\
Fundación para la Conservación y el Desarrollo Sostenible (Colombia) & 1 & Gabon & 948 & Gambia & 180 & Gazprom & 2\\
\addlinespace
\cellcolor{gray!10}{Gbowee Peace Foundation Africa} & \cellcolor{gray!10}{1} & \cellcolor{gray!10}{Geneva Academy Of International Humanitarian Law And Human Rights} & \cellcolor{gray!10}{1} & \cellcolor{gray!10}{Geneva International Centre For Humanitarian Demining} & \cellcolor{gray!10}{3} & \cellcolor{gray!10}{Georgia} & \cellcolor{gray!10}{141}\\
Germany & 1928 & Ghana & 823 & Gisha & 2 & Global Center on Cooperative Security & 1\\
\cellcolor{gray!10}{Global Coalition To Protect Education From Attacks} & \cellcolor{gray!10}{1} & \cellcolor{gray!10}{Global Initiative Against Transnational Organized Crime} & \cellcolor{gray!10}{1} & \cellcolor{gray!10}{Global Vaccine Alliance} & \cellcolor{gray!10}{1} & \cellcolor{gray!10}{Goldman Sachs} & \cellcolor{gray!10}{1}\\
Governing Council Of Iraq & 1 & Grand Imam of Al-Azhar Al-Sharif & 1 & Greece & 338 & Grenada & 1\\
\cellcolor{gray!10}{Gro Intelligence} & \cellcolor{gray!10}{1} & \cellcolor{gray!10}{Group of Five for the Sahel} & \cellcolor{gray!10}{1} & \cellcolor{gray!10}{Group Of Five For The Sahel} & \cellcolor{gray!10}{5} & \cellcolor{gray!10}{Group of Friends of Accountability} & \cellcolor{gray!10}{1}\\
\addlinespace
Group of Friends of Action on Conflict and Hunger & 1 & Group Of Friends Of Children And Armed Conflict & 1 & Group Of Friends Of The African Women Leaders Network & 1 & Group of Friends of the Responsibility to Protect & 2\\
\cellcolor{gray!10}{Group Of Friends Of Women In Afghanistan} & \cellcolor{gray!10}{1} & \cellcolor{gray!10}{Group of Friends on Children and Armed Conflict} & \cellcolor{gray!10}{1} & \cellcolor{gray!10}{Group of Friends on Climate and Security} & \cellcolor{gray!10}{1} & \cellcolor{gray!10}{Group Of Friends On Climate And Security} & \cellcolor{gray!10}{1}\\
Groupe De Recherche, D'etude Et De Formation Femme-Action & 1 & Groupe Lotus And International Federation For Human Rights & 1 & Guatemala & 450 & Guinea & 362\\
\cellcolor{gray!10}{Guinea-Bissau} & \cellcolor{gray!10}{148} & \cellcolor{gray!10}{Gulf Cooperation Council} & \cellcolor{gray!10}{4} & \cellcolor{gray!10}{Gulf Of Guinea Commission} & \cellcolor{gray!10}{4} & \cellcolor{gray!10}{Guyana} & \cellcolor{gray!10}{16}\\
Haiti & 100 & Haiti Liberté & 1 & Haitian Bars Federation & 1 & Haki Africa & 1\\
\addlinespace
\cellcolor{gray!10}{Harvard University (Cambridge, Mass.)} & \cellcolor{gray!10}{1} & \cellcolor{gray!10}{Hashemite Kingdom Of Jordan} & \cellcolor{gray!10}{1} & \cellcolor{gray!10}{High Representative For Bosnia And Herzegovina} & \cellcolor{gray!10}{1} & \cellcolor{gray!10}{High Representative for the Implementation of the Peace Agreement on Bosnia and Herzegovina} & \cellcolor{gray!10}{1}\\
Holy See & 90 & Honduras & 166 & Hope Restoration South Sudan & 1 & Hudson Institute & 1\\
\cellcolor{gray!10}{Human Rights Watch} & \cellcolor{gray!10}{1} & \cellcolor{gray!10}{Hungary} & \cellcolor{gray!10}{332} & \cellcolor{gray!10}{Huquqyat} & \cellcolor{gray!10}{1} & \cellcolor{gray!10}{IAEA} & \cellcolor{gray!10}{17}\\
ICAO & 1 & Iceland & 61 & ICJ & 1 & IMF & 3\\
\cellcolor{gray!10}{IMPACT - Civil Society Research and Development} & \cellcolor{gray!10}{1} & \cellcolor{gray!10}{Independent Anti-Slavery Commissioner} & \cellcolor{gray!10}{1} & \cellcolor{gray!10}{Independent Commission For Overseeing The Implementation Of The Constitution} & \cellcolor{gray!10}{1} & \cellcolor{gray!10}{India} & \cellcolor{gray!10}{1371}\\
\addlinespace
Indonesia & 1208 & Inkatha Freedom Party & 1 & Insight Strategy Partners & 1 & Instancia Especial de Alto Nivel Pueblos Étnicos & 1\\
\cellcolor{gray!10}{Institute For Security Studies} & \cellcolor{gray!10}{1} & \cellcolor{gray!10}{Instituto Rio Branco} & \cellcolor{gray!10}{1} & \cellcolor{gray!10}{Intando Yesizwe Party} & \cellcolor{gray!10}{1} & \cellcolor{gray!10}{Inter-American Development Bank} & \cellcolor{gray!10}{2}\\
Inter-Governmental Action Group Against Money Laundering In West Africa & 1 & Inter-Parliamentary Union & 1 & Intergovernmental Authority On Development & 2 & International Center For Transitional Justice & 1\\
\cellcolor{gray!10}{International Centre For The Study Of Radicalisation And Political Violence} & \cellcolor{gray!10}{2} & \cellcolor{gray!10}{International Committee Of The Red Cross} & \cellcolor{gray!10}{72} & \cellcolor{gray!10}{International Conference On The Great Lakes Region} & \cellcolor{gray!10}{4} & \cellcolor{gray!10}{International Court Of Justice} & \cellcolor{gray!10}{12}\\
International Criminal Court & 36 & International Crisis Group & 3 & International Federation Of Red Cross And Red Crescent Societies & 2 & International Humanitarian Fact-Finding Commission & 1\\
\addlinespace
\cellcolor{gray!10}{International Institute For Justice And The Rule Of Law} & \cellcolor{gray!10}{1} & \cellcolor{gray!10}{International Labour Organization} & \cellcolor{gray!10}{1} & \cellcolor{gray!10}{International Maritime Organization} & \cellcolor{gray!10}{1} & \cellcolor{gray!10}{International Organization for Migration} & \cellcolor{gray!10}{2}\\
International Organization Of La Francophonie & 11 & International Peace Institute & 3 & International Rescue Committee & 2 & International Residual Mechanism For Criminal Tribunals & 16\\
\cellcolor{gray!10}{Interpol} & \cellcolor{gray!10}{12} & \cellcolor{gray!10}{INTERPOL} & \cellcolor{gray!10}{1} & \cellcolor{gray!10}{IOM} & \cellcolor{gray!10}{1} & \cellcolor{gray!10}{Ir Amim} & \cellcolor{gray!10}{1}\\
Iran (Islamic Republic of) & 457 & Iraq & 246 & Iraq Foundation & 1 & Iraqi Al-Amal Association & 1\\
\cellcolor{gray!10}{Iraqi Health and Social Care Organization} & \cellcolor{gray!10}{1} & \cellcolor{gray!10}{Iraqi Women's Network} & \cellcolor{gray!10}{1} & \cellcolor{gray!10}{IRC} & \cellcolor{gray!10}{1} & \cellcolor{gray!10}{Ireland} & \cellcolor{gray!10}{825}\\
\addlinespace
Ireland And Mexico & 1 & Israel & 430 & Italian Ministry Of Culture & 1 & Italy & 922\\
\cellcolor{gray!10}{Jamaica} & \cellcolor{gray!10}{385} & \cellcolor{gray!10}{Japan} & \cellcolor{gray!10}{2055} & \cellcolor{gray!10}{Jerusalem Legal Aid And Human Rights} & \cellcolor{gray!10}{1} & \cellcolor{gray!10}{Jeunesse en Marche pour le Développement en Centrafrique} & \cellcolor{gray!10}{1}\\
Jigsaw & 1 & Joan B. Kroc Institute for International Peace Studies (Notre Dame, Indiana) & 1 & Johns Hopkins University (Baltimore, Md.). School of Advanced International Studies (Washington, D.C.) & 1 & Joint Monitoring And Evaluation Commission & 8\\
\cellcolor{gray!10}{Jordan} & \cellcolor{gray!10}{675} & \cellcolor{gray!10}{Kaladan Press} & \cellcolor{gray!10}{1} & \cellcolor{gray!10}{Kazakhstan} & \cellcolor{gray!10}{585} & \cellcolor{gray!10}{Keeping Children Safe} & \cellcolor{gray!10}{1}\\
Kenya & 659 & King's College London & 1 & Kiribati & 2 & Kofi Annan International Peacekeeping Training Centre & 2\\
\addlinespace
\cellcolor{gray!10}{Kosovo} & \cellcolor{gray!10}{84} & \cellcolor{gray!10}{Kuwait} & \cellcolor{gray!10}{740} & \cellcolor{gray!10}{Kyrgyzstan} & \cellcolor{gray!10}{36} & \cellcolor{gray!10}{La Strada-Ukraine} & \cellcolor{gray!10}{1}\\
Lake Chad Basin Commission & 1 & Lao People's Democratic Republic & 6 & Latvia & 68 & Lawyers For Justice In Libya & 2\\
\cellcolor{gray!10}{League Of Arab States} & \cellcolor{gray!10}{95} & \cellcolor{gray!10}{Lebanon} & \cellcolor{gray!10}{471} & \cellcolor{gray!10}{Lesotho} & \cellcolor{gray!10}{15} & \cellcolor{gray!10}{Liberia} & \cellcolor{gray!10}{63}\\
Liberty And Justice & 1 & Libya & 439 & Libyan Political Dialogue Forum & 1 & Liechtenstein & 217\\
\cellcolor{gray!10}{Liga Internacional de Mujeres por la Paz y Libertad (Colombia)} & \cellcolor{gray!10}{1} & \cellcolor{gray!10}{Lithuania} & \cellcolor{gray!10}{560} & \cellcolor{gray!10}{Lutte Pour Le Changement} & \cellcolor{gray!10}{1} & \cellcolor{gray!10}{Luxembourg} & \cellcolor{gray!10}{388}\\
\addlinespace
Ma'rib Girls Foundation For Development & 1 & Malala Fund & 1 & Malawi & 24 & Malaysia & 914\\
\cellcolor{gray!10}{Maldives} & \cellcolor{gray!10}{70} & \cellcolor{gray!10}{Mali} & \cellcolor{gray!10}{351} & \cellcolor{gray!10}{Mali Musso} & \cellcolor{gray!10}{1} & \cellcolor{gray!10}{Malta} & \cellcolor{gray!10}{449}\\
Marshall Islands & 8 & Mastercard & 1 & Mauritania & 26 & Mauritius & 258\\
\cellcolor{gray!10}{Médecins Du Monde} & \cellcolor{gray!10}{3} & \cellcolor{gray!10}{Medecins Sans Frontieres} & \cellcolor{gray!10}{2} & \cellcolor{gray!10}{Mexico} & \cellcolor{gray!10}{1412} & \cellcolor{gray!10}{Micronesia (Federated States of)} & \cellcolor{gray!10}{2}\\
Microsoft & 1 & Missile Technology Control Regime & 1 & Moby Group & 1 & Monaco & 7\\
\addlinespace
\cellcolor{gray!10}{Mongolia} & \cellcolor{gray!10}{15} & \cellcolor{gray!10}{Montenegro} & \cellcolor{gray!10}{40} & \cellcolor{gray!10}{Morocco} & \cellcolor{gray!10}{586} & \cellcolor{gray!10}{Movement Of Non-Aligned Countries} & \cellcolor{gray!10}{1}\\
Mozambique & 437 & Municipal Association Of Women And Defender Of Afro-Colombian Territorial And Human Rights & 1 & Musahala & 1 & Mwatana Organization For Human Rights & 2\\
\cellcolor{gray!10}{Myanmar} & \cellcolor{gray!10}{75} & \cellcolor{gray!10}{Namibia} & \cellcolor{gray!10}{404} & \cellcolor{gray!10}{National Constitution Amendment Committee} & \cellcolor{gray!10}{2} & \cellcolor{gray!10}{National Council Of Timorese Resistance} & \cellcolor{gray!10}{2}\\
National Independent Electoral Commission Of Somalia & 2 & National People's Party & 1 & NATO & 30 & Nauru & 6\\
\cellcolor{gray!10}{Naweza} & \cellcolor{gray!10}{1} & \cellcolor{gray!10}{Neem Foundation} & \cellcolor{gray!10}{1} & \cellcolor{gray!10}{Nepal} & \cellcolor{gray!10}{98} & \cellcolor{gray!10}{Netherlands} & \cellcolor{gray!10}{867}\\
\addlinespace
Network For Women’s Leadership In The Central African Republic & 1 & Network Of Civil Society Organizations In Borno State & 1 & Network on Peace and Security for Women in the ECOWAS Region & 2 & Networkd of Women-led Organizations of the Lake Chad Basin & 1\\
\cellcolor{gray!10}{New Development Bank} & \cellcolor{gray!10}{1} & \cellcolor{gray!10}{New Zealand} & \cellcolor{gray!10}{838} & \cellcolor{gray!10}{NGO Working Group On Women, Peace And Security} & \cellcolor{gray!10}{15} & \cellcolor{gray!10}{Nicaragua} & \cellcolor{gray!10}{52}\\
Niger & 463 & Nigeria & 1114 & Niue & 1 & North Macedonia & 8\\
\cellcolor{gray!10}{North West Syria NGO Forum} & \cellcolor{gray!10}{1} & \cellcolor{gray!10}{Norway} & \cellcolor{gray!10}{1121} & \cellcolor{gray!10}{Norwegian Refugee Council} & \cellcolor{gray!10}{2} & \cellcolor{gray!10}{Nuclear Suppliers Group} & \cellcolor{gray!10}{1}\\
Office Of The Director Of Public Prosecutions In Kenya & 1 & Oman & 189 & OPCW & 4 & Organisation of Islamic Cooperation & 3\\
\addlinespace
\cellcolor{gray!10}{Organization For Democracy And Economic Development - Guam} & \cellcolor{gray!10}{1} & \cellcolor{gray!10}{Organization For Policy Research And Development Studies} & \cellcolor{gray!10}{2} & \cellcolor{gray!10}{Organization For Responsive Governance} & \cellcolor{gray!10}{1} & \cellcolor{gray!10}{Organization for the Prohibition of Chemical Weapons} & \cellcolor{gray!10}{4}\\
Organization Of American States & 19 & Organization Of Islamic Cooperation & 13 & Organization Of The Islamic Conference & 37 & OSCE & 78\\
\cellcolor{gray!10}{Oxfam} & \cellcolor{gray!10}{2} & \cellcolor{gray!10}{Pacific Community} & \cellcolor{gray!10}{1} & \cellcolor{gray!10}{Pacific Islands Forum} & \cellcolor{gray!10}{1} & \cellcolor{gray!10}{Pacific Small Island Developing States} & \cellcolor{gray!10}{2}\\
Paiman Alumni Trust & 1 & Paix Pour L'enfance & 1 & Pakistan & 1205 & Palau & 5\\
\cellcolor{gray!10}{Palestine} & \cellcolor{gray!10}{248} & \cellcolor{gray!10}{Palestine Red Crescent Society} & \cellcolor{gray!10}{1} & \cellcolor{gray!10}{Palestinian Center For Policy And Survey Research} & \cellcolor{gray!10}{1} & \cellcolor{gray!10}{Pan Africanist Congress Of Azania} & \cellcolor{gray!10}{1}\\
\addlinespace
Panama & 234 & Papua New Guinea & 30 & Paraguay & 25 & Parents Circle & 1\\
\cellcolor{gray!10}{Peace Track Initiative} & \cellcolor{gray!10}{1} & \cellcolor{gray!10}{Peace Track Initiative Yemen} & \cellcolor{gray!10}{1} & \cellcolor{gray!10}{Peru} & \cellcolor{gray!10}{1102} & \cellcolor{gray!10}{Philippines} & \cellcolor{gray!10}{407}\\
Physicians For Human Rights & 2 & Plan International Nigeria & 1 & Plurielles Haiti & 1 & Plurinational State Of Bolivia & 2\\
\cellcolor{gray!10}{Poland} & \cellcolor{gray!10}{1001} & \cellcolor{gray!10}{Policité} & \cellcolor{gray!10}{1} & \cellcolor{gray!10}{Policy And Senior Adviser At Physicians For Human Rights} & \cellcolor{gray!10}{1} & \cellcolor{gray!10}{Portugal} & \cellcolor{gray!10}{688}\\
Programme For Coordination And Assistance For Security And Development (Pcased) & 4 & Project Renew, Norwegian People’s Aid Viet Nam & 1 & Qatar & 408 & Rawadari & 1\\
\addlinespace
\cellcolor{gray!10}{Reconstituted Joint Monitoring And Evaluation Commission} & \cellcolor{gray!10}{2} & \cellcolor{gray!10}{Reconstituted Joint Monitoring and Evaluation Commission (South Sudan)} & \cellcolor{gray!10}{1} & \cellcolor{gray!10}{Rede Feto} & \cellcolor{gray!10}{1} & \cellcolor{gray!10}{Regional Center for Human Rights} & \cellcolor{gray!10}{1}\\
Regional Centre On Small Arms In The Great Lakes Region, The Horn Of Africa And Bordering States & 1 & Release Me & 1 & Reporters Without Borders & 1 & Republic Of Korea & 774\\
\cellcolor{gray!10}{Republic Of Moldova} & \cellcolor{gray!10}{10} & \cellcolor{gray!10}{Reseau Des Femmes Pour La Defense Des Droits Et La Paix} & \cellcolor{gray!10}{1} & \cellcolor{gray!10}{Réseau des jeunes leaders pour la gestion durable des écosystèmes forestiers et humides d'Afrique Centrale} & \cellcolor{gray!10}{1} & \cellcolor{gray!10}{Réseau Ivoirien Pour La Défense Des Droits De L’enfant Et De La Femme} & \cellcolor{gray!10}{1}\\
Rien Sans Les Femmes & 1 & Right To Protection & 1 & Rights And Resources Initiative/Group & 1 & Romania & 371\\
\cellcolor{gray!10}{Rossiya Segodnya} & \cellcolor{gray!10}{1} & \cellcolor{gray!10}{Russian Federation} & \cellcolor{gray!10}{5101} & \cellcolor{gray!10}{Russian Reconciliation Center For Syria} & \cellcolor{gray!10}{2} & \cellcolor{gray!10}{Russkaia pravoslavnaia tserkov} & \cellcolor{gray!10}{2}\\
\addlinespace
Rwanda & 697 & Sahel Alliance & 1 & Saint Kitts And Nevis & 1 & Saint Lucia & 4\\
\cellcolor{gray!10}{Saint Vincent And The Grenadines} & \cellcolor{gray!10}{297} & \cellcolor{gray!10}{Samoa} & \cellcolor{gray!10}{6} & \cellcolor{gray!10}{San Marino} & \cellcolor{gray!10}{11} & \cellcolor{gray!10}{Sana'a Center For Strategic Studies} & \cellcolor{gray!10}{2}\\
Sao Tome And Principe & 3 & Saudi Arabia & 150 & Sauti Ya Mama Mukongomani/Voice Of Congolese Women & 1 & Save Act Mine & 1\\
\cellcolor{gray!10}{Save The Children} & \cellcolor{gray!10}{5} & \cellcolor{gray!10}{Save Ukraine} & \cellcolor{gray!10}{5} & \cellcolor{gray!10}{Sawa For Development And Aid} & \cellcolor{gray!10}{1} & \cellcolor{gray!10}{School of Leadership, Afghanistan} & \cellcolor{gray!10}{1}\\
Search For Common Ground & 2 & Security Council Procedure & 3 & Security Council Report & 8 & Senegal & 510\\
\addlinespace
\cellcolor{gray!10}{Serbia} & \cellcolor{gray!10}{177} & \cellcolor{gray!10}{Serbia And Montenegro} & \cellcolor{gray!10}{31} & \cellcolor{gray!10}{Sesame Workshop} & \cellcolor{gray!10}{1} & \cellcolor{gray!10}{Seychelles} & \cellcolor{gray!10}{3}\\
Shanghai Cooperation Organization & 3 & Sheba Youth Foundation For Development & 1 & Siemens & 2 & Sierra Leone & 122\\
\cellcolor{gray!10}{Similar Ground} & \cellcolor{gray!10}{1} & \cellcolor{gray!10}{Singapore} & \cellcolor{gray!10}{416} & \cellcolor{gray!10}{Slovak Republic} & \cellcolor{gray!10}{334} & \cellcolor{gray!10}{Slovenia} & \cellcolor{gray!10}{308}\\
Small Arms Survey & 1 & Solidarité Féminine pour la paix et le Développement intégral & 1 & Solidarity Party & 1 & Solomon Islands & 14\\
\cellcolor{gray!10}{Solutions For Sustainable Society} & \cellcolor{gray!10}{1} & \cellcolor{gray!10}{Somali National Women’s Organization} & \cellcolor{gray!10}{1} & \cellcolor{gray!10}{Somali Women Development Centre} & \cellcolor{gray!10}{1} & \cellcolor{gray!10}{Somali Women's Leadership Initiative} & \cellcolor{gray!10}{1}\\
\addlinespace
Somali Women’s Studies Centre & 1 & Somalia & 130 & Somalia Youth Development Network & 1 & Sos Center For Youth Capabilities Development & 1\\
\cellcolor{gray!10}{South Africa} & \cellcolor{gray!10}{1347} & \cellcolor{gray!10}{South African Communist Party} & \cellcolor{gray!10}{1} & \cellcolor{gray!10}{South Sudan} & \cellcolor{gray!10}{109} & \cellcolor{gray!10}{South Sudan Women With Disabilities Network} & \cellcolor{gray!10}{1}\\
South Sudan Women's Empowerment Network & 1 & South Sudan Women's Monthly Forum & 1 & Sovereign Order of Malta & 5 & Spain & 1352\\
\cellcolor{gray!10}{Sri Lanka} & \cellcolor{gray!10}{102} & \cellcolor{gray!10}{Stimson Center} & \cellcolor{gray!10}{2} & \cellcolor{gray!10}{Strategic Foresight Group} & \cellcolor{gray!10}{1} & \cellcolor{gray!10}{Strategic Initiative For Women In The Horn Of Africa} & \cellcolor{gray!10}{2}\\
Sudan & 319 & Sudan People's Liberation Movement & 4 & Sudan Social Development Organization & 1 & Suriname & 3\\
\addlinespace
\cellcolor{gray!10}{Sustainable Pacific Consultancy Niue} & \cellcolor{gray!10}{1} & \cellcolor{gray!10}{Swaziland} & \cellcolor{gray!10}{6} & \cellcolor{gray!10}{Sweden} & \cellcolor{gray!10}{889} & \cellcolor{gray!10}{Switzerland} & \cellcolor{gray!10}{596}\\
Synergy Of Women For Victims Of Sexual Violence & 1 & Syria Bright Future & 1 & Syria Relief And Development & 2 & Syrian American Medical Society & 3\\
\cellcolor{gray!10}{Syrian Arab Red Crescent} & \cellcolor{gray!10}{2} & \cellcolor{gray!10}{Syrian Arab Republic} & \cellcolor{gray!10}{712} & \cellcolor{gray!10}{Syrian Center For Media And Freedom Of Expression} & \cellcolor{gray!10}{1} & \cellcolor{gray!10}{Syrian Emergency Task Force} & \cellcolor{gray!10}{1}\\
Syrian National Conference & 1 & Syrian Women's League & 1 & Syrian Women’s Political Movement & 2 & Tajikistan & 37\\
\cellcolor{gray!10}{Tamazight Women's Movement} & \cellcolor{gray!10}{3} & \cellcolor{gray!10}{Thailand} & \cellcolor{gray!10}{137} & \cellcolor{gray!10}{The Elders} & \cellcolor{gray!10}{4} & \cellcolor{gray!10}{The Grayzone (Firm)} & \cellcolor{gray!10}{1}\\
\addlinespace
The Liaison Office & 1 & Timor-Leste & 49 & Together We Build It & 2 & Togo & 319\\
\cellcolor{gray!10}{Tonga} & \cellcolor{gray!10}{7} & \cellcolor{gray!10}{Trial International} & \cellcolor{gray!10}{2} & \cellcolor{gray!10}{Trinidad And Tobago} & \cellcolor{gray!10}{21} & \cellcolor{gray!10}{Tunisia} & \cellcolor{gray!10}{592}\\
Turkey & 732 & Türkiye & 61 & Turkmenistan & 5 & Tuvalu & 7\\
\cellcolor{gray!10}{U.S./Middle East Project} & \cellcolor{gray!10}{3} & \cellcolor{gray!10}{Uganda} & \cellcolor{gray!10}{529} & \cellcolor{gray!10}{Ukraine} & \cellcolor{gray!10}{983} & \cellcolor{gray!10}{Ukrainian Trade Union of Law Workers} & \cellcolor{gray!10}{2}\\
Ukrainian Women's Fund & 1 & UN & 4709 & UN High Commissioner for Human Rights & 2 & UN High Commissioner for Refugees & 2\\
\addlinespace
\cellcolor{gray!10}{UN Institute for Disarmament Research} & \cellcolor{gray!10}{1} & \cellcolor{gray!10}{UN Interim Force in Lebanon} & \cellcolor{gray!10}{2} & \cellcolor{gray!10}{UN Mission in South Sudan} & \cellcolor{gray!10}{2} & \cellcolor{gray!10}{UN Multidimensional Integrated Stabilization Mission in the Central African Republic} & \cellcolor{gray!10}{2}\\
UN Office on Drugs and Crime & 3 & UN Organization Stabilization Mission in the Democratic Republic of the Congo & 1 & UN-Women & 7 & UNFPA & 1\\
\cellcolor{gray!10}{UNHCR} & \cellcolor{gray!10}{1} & \cellcolor{gray!10}{UNICEF} & \cellcolor{gray!10}{4} & \cellcolor{gray!10}{United Arab Emirates} & \cellcolor{gray!10}{944} & \cellcolor{gray!10}{United Democratic Movement} & \cellcolor{gray!10}{1}\\
United Kingdom Of Great Britain And Northern Ireland & 5518 & United Nations Office On Drugs And Crime & 1 & United Nations University & 1 & United Republic Of Tanzania & 236\\
\cellcolor{gray!10}{United States Of America} & \cellcolor{gray!10}{5484} & \cellcolor{gray!10}{University of Oxford} & \cellcolor{gray!10}{1} & \cellcolor{gray!10}{UNRWA} & \cellcolor{gray!10}{2} & \cellcolor{gray!10}{URU} & \cellcolor{gray!10}{2}\\
\addlinespace
Uruguay & 752 & Uzbekistan & 35 & Vanuatu & 1 & Venezuela (Bolivarian Republic Of) & 680\\
\cellcolor{gray!10}{Viet Nam} & \cellcolor{gray!10}{976} & \cellcolor{gray!10}{Viva La Vida} & \cellcolor{gray!10}{1} & \cellcolor{gray!10}{Voice Amplified} & \cellcolor{gray!10}{1} & \cellcolor{gray!10}{Voice Of Libyan Women} & \cellcolor{gray!10}{1}\\
Volontariat Pour Le Développement D’haïti & 1 & WASHM Center for Women's Studies & 1 & Watchlist On Children And Armed Conflict & 4 & West African Action Network On Small Arms & 1\\
\cellcolor{gray!10}{WHO} & \cellcolor{gray!10}{2} & \cellcolor{gray!10}{Women And Children Legal Research Foundation} & \cellcolor{gray!10}{1} & \cellcolor{gray!10}{Women And Peace Studies Organization} & \cellcolor{gray!10}{2} & \cellcolor{gray!10}{Women Empowerment Organization} & \cellcolor{gray!10}{1}\\
Women For Women International & 2 & Women in Action for Women & 1 & Women In Law And Development In Africa & 1 & Women In Peacebuilding Network & 1\\
\addlinespace
\cellcolor{gray!10}{Women Now for Development (Syrian Arab Republic)} & \cellcolor{gray!10}{1} & \cellcolor{gray!10}{Women's Centre For Legal Aid And Counselling} & \cellcolor{gray!10}{1} & \cellcolor{gray!10}{Women's Empowerment Network} & \cellcolor{gray!10}{1} & \cellcolor{gray!10}{Women's Initiatives For Gender Justice} & \cellcolor{gray!10}{1}\\
Women's Institute for Alternative Development & 1 & Women's International League For Peace And Freedom & 1 & Women's League of Burma & 1 & Women's Monthly Forum On Peace And Political Processes In South Sudan & 1\\
\cellcolor{gray!10}{Women's Refugee Route} & \cellcolor{gray!10}{1} & \cellcolor{gray!10}{Women's Solidarity For Peace And Integral Development} & \cellcolor{gray!10}{1} & \cellcolor{gray!10}{World Bank} & \cellcolor{gray!10}{30} & \cellcolor{gray!10}{World Customs Organization} & \cellcolor{gray!10}{1}\\
World Food Programme & 3 & World Health Organization & 2 & World Meteorological Organization & 1 & World Wildlife And One Young World & 1\\
\cellcolor{gray!10}{World Young Women Christian Association} & \cellcolor{gray!10}{1} & \cellcolor{gray!10}{Yale University (New Haven, Conn.)} & \cellcolor{gray!10}{2} & \cellcolor{gray!10}{Yemen} & \cellcolor{gray!10}{143} & \cellcolor{gray!10}{Youth Association For Active Citizenship And Democracy} & \cellcolor{gray!10}{1}\\
\addlinespace
Youth Initiative For Human Rights In Bosnia And Herzegovina & 1 & Yugoslavia & 54 & Zaire & 5 & Zambia & 39\\
\cellcolor{gray!10}{Zan Times} & \cellcolor{gray!10}{1} & \cellcolor{gray!10}{Zanele Mbeki Development Trust} & \cellcolor{gray!10}{1} & \cellcolor{gray!10}{Zanmi Lasante} & \cellcolor{gray!10}{2} & \cellcolor{gray!10}{Zimbabwe} & \cellcolor{gray!10}{99}\\
Zone Of Peace And Cooperation Of The South Atlantic & 1 &  &  &  &  &  & \\*
\end{longtable}
\endgroup{}

%% file: main.bib
@article{eckhard2023international,
  title={International bureaucrats in the UN Security Council debates: A speaker-topic network analysis},
  author={Eckhard, Steffen and Patz, Ronny and Sch{\"o}nfeld, Mirco and van Meegdenburg, Hilde},
  journal={Journal of European public policy},
  volume={30},
  number={2},
  pages={214--233},
  year={2023},
  publisher={Taylor \& Francis}
}

@book{2023jankaus,
    author = {Jankauskas, Vytautas and Eckhard, Steffen},
    title = {The Politics of Evaluation in International Organizations},
    publisher = {Oxford University Press},
    year = {2023},
    month = {05},
    isbn = {9780192855206},
    doi = {10.1093/oso/9780192855206.001.0001},
    url = {https://doi.org/10.1093/oso/9780192855206.001.0001},
}

@article{Jankin2024,
  title = {Words to unite nations: The complete United Nations General Debate Corpus,  1946–present},
  ISSN = {1460-3578},
  DOI = {10.1177/00223433241275335},
  journal = {Journal of Peace Research},
  publisher = {SAGE Publications},
  author = {Jankin,  Slava and Baturo,  Alexander and Dasandi,  Niheer},
  year = {2024},
  month = nov 
}

@inproceedings{rodven-eide-etal-2023-unsc,
    title = "The {UNSC}-Graph: An Extensible Knowledge Graph for the {UNSC} Corpus",
    author = "R{\o}dven-Eide, Stian  and
      Zaczynska, Karolina  and
      Pires, Antonio  and
      Patz, Ronny  and
      Stede, Manfred",
    editor = "Klamm, Christopher  and
      Lapesa, Gabriella  and
      Gold, Valentin  and
      Gessler, Theresa  and
      Ponzetto, Simone Paolo",
    booktitle = "Proceedings of the 3rd Workshop on Computational Linguistics for the Political and Social Sciences",
    month = sep,
    year = "2023",
    address = "Ingolstadt, Germany",
    publisher = "Association for Computational Lingustics",
    url = "https://aclanthology.org/2023.cpss-1.7",
    pages = "69--74",
}

@article{Mesquita2024,
  title = {The references of the nations: Introducing a corpus of United Nations General Assembly resolutions since 1946 and their citation network},
  ISSN = {1460-3578},
  DOI = {10.1177/00223433241254997},
  journal = {Journal of Peace Research},
  publisher = {SAGE Publications},
  author = {Mesquita,  Rafael and Pires,  Antonio},
  year = {2024},
  month = aug 
}

@article{Scherzinger2022,
  title = {Unbowed,  unbent,  unbroken? Examining the validity of the responsibility to protect},
  volume = {58},
  ISSN = {1460-3691},
  DOI = {10.1177/00108367221093155},
  number = {1},
  journal = {Cooperation and Conflict},
  publisher = {SAGE Publications},
  author = {Scherzinger,  Johannes},
  year = {2022},
  month = may,
  pages = {81–101}
}

@article{Verbeek2024,
  title = {‘Wolf Warriors’ in the {UN Security Council}? Investigating power shifts through blaming},
  volume = {15},
  ISSN = {1758-5899},
  DOI = {10.1111/1758-5899.13372},
  number = {S2},
  journal = {Global Policy},
  publisher = {Wiley},
  author = {Verbeek,  Nicolas},
  year = {2024},
  month = may,
  pages = {38–50}
}

@article{turco2024speaking,
  title={Speaking Volumes: Introducing the UNGA Speech Corpus},
  author={Turco, Linnea R},
  journal={International Studies Quarterly},
  volume={68},
  number={1},
  pages={sqae001},
  year={2024},
  publisher={Oxford University Press}
}

@article{Sakamoto2023,
  title = {Threat Conceptions in Global Security Discourse: Analyzing the Speech Records of the United Nations Security Council,  1990–2019},
  volume = {67},
  ISSN = {1468-2478},
  url = {http://dx.doi.org/10.1093/isq/sqad067},
  DOI = {10.1093/isq/sqad067},
  number = {3},
  journal = {International Studies Quarterly},
  publisher = {Oxford University Press (OUP)},
  author = {Sakamoto,  Takuto},
  year = {2023},
  month = jun 
}

@incollection{2022-eckhard-politics,
  title={The politics of evaluation in international organizations},
  author={Eckhard, Steffen and Jankauskas, Vytautas},
  booktitle={International Public Administrations in Global Public Policy},
  pages={183--198},
  year={2022},
  publisher={Routledge}
}

@book{2024-reinalda-handb-io,
  title={Routledge handbook of international organization},
  author={Reinalda, Bob},
  year={2024},
  publisher={Routledge London}
}

@book{2019-rittb-io,
  title={International organization},
  author={Rittberger, Volker and Zangl, Bernhard and Kruck, Andreas and Dijkstra, Hylke},
  year={2019},
  publisher={Bloomsbury Publishing}
}

@book{2022-knill-ipa,
  title = {International Public Administrations in Global Public Policy: Sources and Effects of Bureaucratic Influence},
  ISBN = {9781003323297},
  url = {http://dx.doi.org/10.4324/9781003323297},
  DOI = {10.4324/9781003323297},
  publisher = {Routledge},
  author = {Knill,  Christoph and Steinebach,  Yves},
  year = {2022},
  month = nov 
}

@article{2023-eckhard-performance,
  title = {The performance of international organizations: a new measure and dataset based on computational text analysis of evaluation reports},
  volume = {18},
  ISSN = {1559-744X},
  DOI = {10.1007/s11558-023-09489-1},
  number = {4},
  journal = {The Review of International Organizations},
  publisher = {Springer Science and Business Media LLC},
  author = {Eckhard,  Steffen and Jankauskas,  Vytautas and Leuschner,  Elena and Burton,  Ian and Kerl,  Tilman and Sevastjanova,  Rita},
  year = {2023},
  month = may,
  pages = {753–776}
}

@article{2021-svanh-intbur,
author = {Svanhildur Thorvaldsdottir and Ronny Patz and Steffen Eckhard},
title ={International bureaucracy and the United Nations system: introduction},
journal = {International Review of Administrative Sciences},
volume = {87},
number = {4},
pages = {695-700},
year = {2021},
doi = {10.1177/00208523211038730}
}

@book{rulesworld,
  year = {2004},
  publisher = {Cornell University Press},
  author = {Michael Barnett and Martha Finnemore},
  title = {Rules for the World}
}

@book{mgnrsglblchange,
  year = {2014},
  publisher = {MIT Press},
  editor = {Frank Biermann and Bernd Siebenhüner},
  title = {Managers of Global Change}
}

@book{trondalunpackingio,
  year = {2014},
  publisher = {Manchester University Press},
  author = {Trondal, Jarle and Marcussen, Martin and Larsson, Torbjorn and Veggeland, Frode},
  title = {Unpacking International Organisations}
}

@Article{Hooghe2015,
author="Hooghe, Liesbet
and Marks, Gary",
title="Delegation and pooling in international organizations",
journal="The Review of International Organizations",
year="2015",
month="9",
day="01",
volume="10",
number="3",
pages="305--328",
issn="1559-744X",
doi="10.1007/s11558-014-9194-4",
}

@book{2006hawkins,
  doi = {10.1017/cbo9780511491368},
  year = {2006},
  publisher = {Cambridge University Press},
  editor = {Hawkins, Darren G. and Lake, David A. and Nielson, Daniel L. and Tierney, Michael J.},
  title = {Delegation and Agency in International Organizations}
}

@article{2018eckhardreform,
author = {Eckhard, Steffen and Patz, Ronny and Schmidt, Sylvia},
title = {Reform efforts, synchronization failure, and international bureaucracy: the case of the {UNESCO} budget crisis},
journal = {Journal of European Public Policy},
volume = {0},
number = {0},
pages = {1-18},
year  = {2018},
publisher = {Routledge},
doi = {10.1080/13501763.2018.1539116}
}

@article{2016eckhardinflburea,
author = {Eckhard, Steffen and Ege, J\"orn},
title = {International bureaucracies and their influence on policy-making: a review of empirical evidence},
journal = {Journal of European Public Policy},
volume = {23},
number = {7},
pages = {960-978},
year  = {2016},
publisher = {Routledge},
doi = {10.1080/13501763.2016.1162837}
}

@book{2017intburea,
  doi = {10.1057/978-1-349-94977-9},
  year = {2017},
  publisher = {Palgrave Macmillan {UK}},
  editor = {Michael W. Bauer and Christoph Knill and Steffen Eckhard},
  title = {International Bureaucracy}
}

@book{abbott_genschel_snidal_zangl_2015,
  doi = {10.1017/cbo9781139979696},
  year = {2014},
  publisher = {Cambridge University Press},
  editor = {Kenneth W. Abbott and Philipp Genschel and Duncan Snidal and Bernhard Zangl},
  title = {International Organizations as Orchestrators}
}

@book{2018-froehlich-williams,
  title={The {UN} {Secretary-General} and the {Security Council}: A Dynamic Relationship},
  author={Fr{\"o}hlich, Manuel and Williams, Abiodun},
  year={2018},
  publisher={Oxford University Press}
}

@Article{quanteda,
  title = {quanteda: An {R} package for the quantitative analysis of textual data},
  journal = {Journal of Open Source Software},
  author = {Kenneth Benoit and Kohei Watanabe and Haiyan Wang and Paul Nulty and Adam Obeng and Stefan Müller and Akitaka Matsuo},
  doi = {10.21105/joss.00774},
  url = {https://quanteda.io},
  volume = {3},
  number = {30},
  pages = {774},
  year = {2018},
}

@inproceedings{2009_smith_tesseract,
 author = {Smith, Ray and Antonova, Daria and Lee, Dar-Shyang},
 title = {Adapting the Tesseract Open Source {OCR} Engine for Multilingual {OCR}},
 booktitle = {Proceedings of the International Workshop on Multilingual {OCR}},
 series = {MOCR '09},
 year = {2009},
 isbn = {978-1-60558-698-4},
 location = {Barcelona, Spain},
 pages = {1:1--1:8},
 articleno = {1},
 numpages = {8},
 doi = {10.1145/1577802.1577804},
 acmid = {1577804},
 publisher = {ACM},
 address = {New York, NY, USA},
}

@INPROCEEDINGS{2007_smith_tesseract,
author={R. {Smith}},
booktitle={Ninth International Conference on Document Analysis and Recognition (ICDAR 2007)},
title={An Overview of the Tesseract {OCR} Engine},
year={2007},
volume={2},
pages={629-633},
doi={10.1109/ICDAR.2007.4376991},
ISSN={1520-5363},
month={9},}

@article{2019-squatrito-shaming_by_io,
author = {Theresa Squatrito and Magnus Lundgren and Thomas Sommerer},
title ={Shaming by international organizations: Mapping condemnatory speech acts across 27 international organizations, 1980–2015},
journal = {Cooperation and Conflict},
year = {2019},
doi = {10.1177/0010836719832339}
}

@misc{2019_unsc_corpus,
author = {Schoenfeld, Mirco and Eckhard, Steffen and Patz, Ronny and Meegdenburg, Hilde van},
publisher = {Harvard Dataverse},
title = "{The UN Security Council debates}",
UNF = {UNF:6:yAQ+0vBM7KhG9ZdyzLJOiA==},
year = {2019},
version = {V1},
doi = {10.7910/DVN/KGVSYH},
}

@book{2014-sievers-unsc_procedure,
  title={The procedure of the {UN Security Council}},
  author={Sievers, Loraine and Daws, Sam},
  year={2014},
  publisher={Oxford University Press},
  address={Oxford}
}

@misc{2018_unsc_discurs,
Author = {Mirco Schoenfeld and Steffen Eckhard and Ronny Patz and Hilde van Meegdenburg},
Title = {Discursive Landscapes and Unsupervised Topic Modeling in {IR}: A Validation of Text-As-Data Approaches through a New Corpus of {UN Security Council} Speeches on Afghanistan},
Year = {2018},
Eprint = {arXiv:1810.05572},
}

@misc{2018_unscafg,
author = {Schoenfeld, Mirco and Eckhard, Steffen and Patz, Ronny and Meegdenburg, Hilde van},
publisher = {Harvard Dataverse},
title = {The {UN Security Council} debates on {Afghanistan}},
year = {2018},
doi = {10.7910/DVN/OM9RG8},
}

@INPROCEEDINGS{2017-pomeroy-un_space,
author={C. {Pomeroy}},
booktitle={2017 International Conference on the Frontiers and Advances in Data Science (FADS)},
title={spaceTexts: A new corpus of speeches in the {UN} committee on the peaceful uses of outer space},
year={2017},
volume={},
number={},
pages={41-46},
doi={10.1109/FADS.2017.8253191},
ISSN={},
month={10},}

@article{2003-king-inf_extract, title={An Automated Information Extraction Tool for International Conflict Data with Performance as Good as Human Coders: A Rare Events Evaluation Design}, volume={57}, DOI={10.1017/S0020818303573064}, number={3}, journal={International Organization}, publisher={Cambridge University Press}, author={King, Gary and Lowe, Will}, year={2003}, pages={617–642}}

@book{2009-bosco-five,
  title={Five to rule them all: the {UN Security Council} and the making of the modern world},
  author={Bosco, David L},
  year={2009},
  publisher={Oxford University Press},
  address={Oxford}
}

@article{2017-baturo-ungd_corpus,
author = {Alexander Baturo and Niheer Dasandi and Slava J. Mikhaylov},
title ={Understanding state preferences with text as data: Introducing the {UN General Debate} corpus},
journal = {Research \& Politics},
volume = {4},
number = {2},
pages = {2053168017712821},
year = {2017},
doi = {10.1177/2053168017712821}
}
